\documentclass[apj]{emulateapj}

\usepackage{graphicx}
\usepackage{float}
\usepackage{color}
\usepackage{CJKutf8}
\usepackage[dvipsnames]{xcolor}
\usepackage[colorlinks=true,citecolor=blue,linkcolor=blue]{hyperref}
\usepackage{dashrule}
\usepackage{subfigure}

\begin{document}

\newcommand {\dem}{DEM\,L\,71\ }
\expandafter\def\csname E0102 \endcsname{E0102}
\newcommand {\E}{\csname E0102 \endcsname} 

\newcommand {\iib}{Type IIb\ }
\newcommand {\xray}{X-ray\ }
\newcommand {\ha}{H$\alpha$\ }
\newcommand {\hb}{H$\beta$\ }
\newcommand {\nii}{[\ion{N}{2}]\ }
\newcommand {\hii}{\ion{H}{2}\ }
\newcommand {\oiii}{[\ion{O}{3}]\ }
\newcommand {\sii}{[\ion{S}{2}]\ }
\newcommand {\kms}{km~s$^{-1}$}
\newcommand {\msun}{$M_\odot$}
\newcommand {\lsun}{$L_\sun$}
\newcommand {\teff}{$T_{eff}$}
\newcommand {\s}{$\sim$}
\newcommand {\hst}{HST\ }
\newcommand {\chandra}{Chandra\ }
\newcommand {\gaia}{Gaia\ }


\title{Searching for Surviving Companion in the Young SMC Supernova Remnant 1E 0102.2-7219}

\author{Chuan-Jui Li \begin{CJK}{UTF8}{bsmi}(李傳睿)\end{CJK}$^{1}$, 
Ivo R. Seitenzahl$^{2}$, Ryoko Ishioka$^1$,\\
You-Hua Chu \begin{CJK}{UTF8}{bsmi}(朱有花)\end{CJK}$^{1,3}$, 
Ashley J. Ruiter$^{2}$,
Fr\'ed\'eric P. A. Vogt$^{4}$}
\affil{$^1$Institute of Astronomy and Astrophysics, Academia Sinica, No.1, Sec. 4, Roosevelt Rd., Taipei 10617, Taiwan\ 
\\ cjli@asiaa.sinica.edu.tw\\
$^2$School of Science, University of New South Wales, Australian Defence Force Academy\\
Canberra, ACT 2600, Australia\\
$^3$Department of Astronomy, University of Illinois at Urbana-Champaign, 1002 West Green Street, \\
Urbana,IL 61801, U.S.A.\\
$^4$European Southern Observatory, Av. Alonso de Córdova 3107, 763 0355 Vitacura, Santiago, Chile}




%
\begin{abstract} 
1E 0102.2--7219 (hereafter E0102) is a young supernova remnant (SNR) in the Small Magellanic Cloud (SMC).  It contains oxygen-rich SN ejecta, a possible neutron star (NS), and a small amount of fast-moving H-rich ejecta material.  These properties are also seen in Cas A, it has thus been suggested that E0102 is also a Type IIb SNR, whose SN progenitor's hydrogen envelope was stripped off possibly via interactions with a companion star. To search for a surviving companion of E0102's SN progenitor, we have used archival Hubble Space Telescope (HST) continuum images to make photometric measurements of stars projected in the SNR to construct color-magnitude diagrams and compare the stars with those expected from surviving companions of Type IIb SNe. We have also used the Multi-Unit Spectroscopic Explorer observations taken with the Very Large Telescope to perform spectroscopic analyses of stars and search for peculiar radial velocities as diagnostics of surviving companions. We further use the HST and Gaia data to inspect proper motions of stars for complementary kinetic studies. No plausible companion candidates are found if the SN explosion site was near the NS, while the B3\,V star 34a may be a plausible companion candidate if the SN explosion site is near the SN ejecta's expansion center.  
If the NS is real and associated with E0102, it needs a $\sim$1000 km s$^{-1}$ runaway velocity, which has been observed in other SNRs and can be acquired from an asymmetric SN explosion or a kick by the SN explosion of a massive companion. 

\end{abstract}

\subjectheadings{ISM: supernova remnants --- ISM: individual objects (1E 0102.2-7219) --- Magellanic Clouds --- stars: massive}

\section{Introduction}  \label{sec:intro}

It has been observed that the majority of massive stars are in binaries (\citealt{Kobulnicky2007, Chini2012, Sana2012, Kobulnicky2014, Moe2017}; see \citealt{Duchene2013} for a review). At the end of a massive star's evolution, the stellar core collapses to form a neutron star (NS) or a black hole (BH), depending on the mass of the star \citep{Sukhbold2018}. If the collapsing core manages to explode, a so-called core-collapse supernova (CCSN) emerges. It is conceivable that a significant fraction of the CCSNe originate from massive stars in binary systems \citep{Kochanek2009, Smith2011, Eldridge2013}.  
Indeed CCSNe have been observed to exhibit a wide variety of spectral characteristics, and some of them have been suggested to be caused by the progenitors' binary evolution.

SNe are broadly divided into Type I and Type II based on the absence and presence of hydrogen lines in their optical spectra, respectively, and each type is further divided into subtypes depending on strengths of helium and silicon lines \citep{Filippenko1997}.  
Among these subtypes, \iib SNe show weak hydrogen lines at early times and these lines fade away over weeks to months \citep{Filippenko1988, Filippenko1993, Filippenko1997, Gal-Yam2016}. After the disappearance of hydrogen lines, the spectra of \iib SNe become similar to those of Type Ib SNe \citep{Barbon1995}. 
\iib SN progenitors have been suggested to be mergers of two stars to produce a very thin layer of hydrogen \citep{Nomoto1995} or simply Wolf-Rayet (WR) stars with hydrogen-poor atmospheres \citep{Georgy2012}; however, a more commonly accepted cause of \iib SNe's deficiency in hydrogen is that their progenitors' hydrogen-rich envelopes have been stripped through interactions with binary companions \citep{Nomoto1993, Woosley1994, Podsiadlowski1993, Stancliffe2009, Claeys2011, Yoon2017}.

To search for companions, photometric measurements of \iib SN progenitors in pre-explosion images were made and modeled, e.g., SNe 1993J \citep{VanDyk2002, Maund2004, Stancliffe2009, Fox2014}, 2008ax \citep{Crockett2008, Folatelli2015}, 2011dh \citep{Maund2011, VanDyk2011, Folatelli2014, Maund2015}, 2013df \citep{VanDyk2014}, and 2016gkg \citep{Kilpatrick2017, Sravan2018} or alternatively post-explosion images (after SNe had faded) were used to directly search for SN progenitors' companions, e.g., SN 2001ig \citep{Ryder2006, Ryder2018}. In general, the conclusion of binary progenitors for \iib SNe is favored over other scenarios. 

Searches for companions of SN progenitors were also made in SN remnants (SNRs) whose SN types had been unambiguously identified using spectra of light echos of their SNe, for example, the \iib SN of Cas A \citep{Krause2008}.
Searches in Cas A have not found a surviving companion \citep{Kochanek2018, Kerzendorf2019, Fraser2019}.  Searches for surviving companions of SN progenitors can be extended to SNRs in the Large and Small Magellanic Clouds (LMC and SMC), such as the LMC Balmer-dominated young Type Ia SNRs  (e.g., \citealt{Li2017, Litke2017, Li2019}). We now report a similar search for the SNR 1E 0102.2--7219 (hereafter \E), a likely \iib SNR in the SMC at a distance of 62 kpc \citep{Graczyk2014, Scowcroft2016}, where 1\farcs0 corresponds to $\sim$ 0.3 pc. 

\begin{figure}[h]  
\begin{center}
\includegraphics[width=21.5pc]{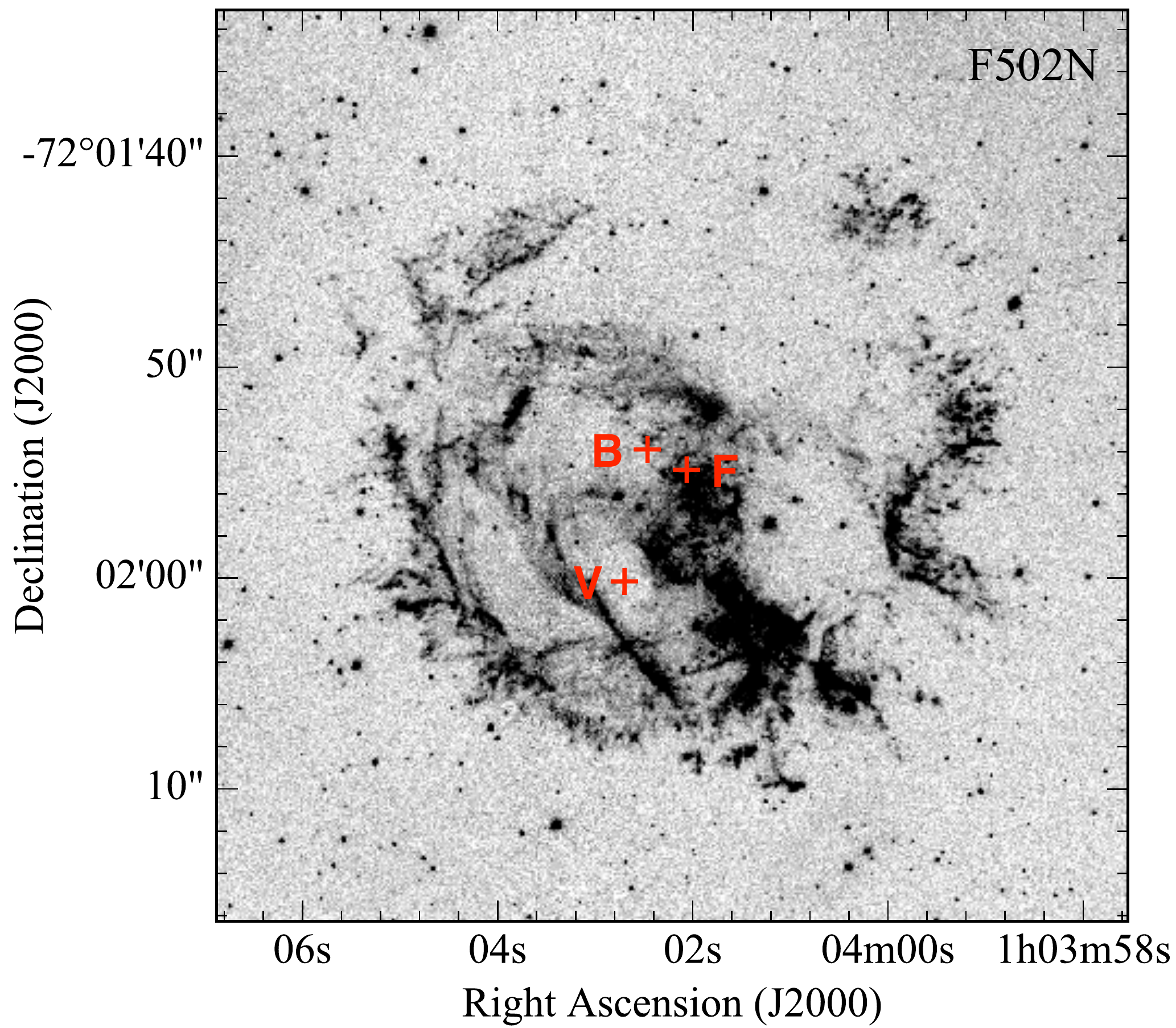}
\caption{F502N image of \E\ obtained with the HST WFC3. The SN explosion sites suggested by \citet{Finkelstein2006}, \citet{Vogt2018}, and \citet{Banovetz2021} are marked by crosses and labeled by F, V, and B, respectively.}
\label{figure:E0102_F502N}
\end{center}
\end{figure}

The SNR \E, shown in Figure \ref{figure:E0102_F502N}, was first identified as such by its bright X-ray luminosity \citep{Seward1981} and found to be oxygen-rich by its high [O III]/H$\alpha$ ratio and optical and UV spectral analyses  \citep{Dopita1981, Blair1989}. Its age can be derived from the radial expansion of the SN ejecta measured from Hubble Space Telescope (HST) images taken at two epochs.  Over a time span of $\sim$8.5 yr, the age was determined to be 2050$\pm$600 yr \citep{Finkelstein2006}, and over a time span of $\sim$19 yr, the age was revised to 1738$\pm$175 yr \citep{Banovetz2021}.
Its initial Type II classification was based on its oxygen-rich SN ejecta
\citep{Tuohy1983}, but was later revised to be Type Ib because of a lack of oxygen-burning products, such as S, Ca, Ar, etc \citep{Blair2000}. Subsequently, \citet{Chevalier2005} considered the SNR's interaction with the SN progenitor's dense wind and suggested a SN Type IIn or IIb.  Recently, \citet{Seitenzahl2018} detected high-velocity Balmer emission within \E, in favor of its \iib classification. The presence of fast-moving hydrogen indicates that the hydrogen envelope of \E's SN progenitor was mostly stripped before the explosion, possibly via interactions with a binary companion star. 

\E\ and the confirmed \iib SNR Cas A share many similar properties. Cas A is found to be oxygen-rich from optical spectra \citep{Kirshner1977, Chevalier1979}, so is \E\ \citep{Dopita1981, Blair1989}. Near the geometric center of Cas A exists an X-ray point source that is likely the NS produced by the SN of Cas A \citep{Tananbaum1999,Chakrabarty2001}, and \E\ may also have a NS within its boundary albeit at an off-center position  \citep{Vogt2018}. While the NS in \E\ is confirmed by \citet{Hebbar2020}, it is also challenged by \citet{Long2020}, who suggest that the X-ray source is an ejecta knot rather than a NS. Within Cas A, fast hydrogen-rich knots were detected to have velocities $\sim$ 6000 \kms, implying condensations from the progenitor’s outer layers \citep{Fesen1988, Fesen1991}, and in \E\ a small amount of fast-moving hydrogen-rich ejecta material is observed to have velocities up to 1785 \kms\ \citep{Seitenzahl2018}. Using the \oiii filaments in \E, \citet{Vogt2010} reconstructed its full three-dimensional structure and suggested a Cas A-like asymmetric bipolar structure, which could be caused by the rotation and collapse of the SN's progenitor star or the disruption of the initial binary system if the SN progenitor star had a companion star.


\E\ is therefore an intriguing target to search for a surviving companion.
We have used HST images to study the physical structure of the SNR as well as the underlying stellar population. 
We have also used the Multi Unit Spectroscopic Explorer (MUSE) observations at the Very Large Telescope (VLT) to carry out spectroscopic analyses of stars to use peculiar radial velocities as diagnostics of surviving companions. Proper motion data from Gaia Early Data Release 3 \citep[EDR3;][]{Gaia2020} are also used to complement the stellar kinematics considerations.  This paper reports our investigation and is organized as follows. Section 2 describes the observations we used. Section 3 details our methodology, and section 4 describes the results. Section 5 discusses implications of the results on the SN progenitor, and a summary is given in section 6.


\section{Observations and Data Reduction} \label{sec:obs}  

\subsection{Hubble Space Telescope Observations}

We use archival HST narrow-band (e.g., F502N) and continuum-band (e.g., F475W, F550M, and F775W) images of SNR \E\ to investigate the nature of its SN progenitor. The narrow-band images are used to examine physical structures of the SNR, while the continuum-band images are used to analyze the underlying stellar populations. 

The available HST images are listed in Table \ref{table:hst}, where the PI, program ID, filter, instrument, date, and exposure time of each observation are given.  Three cameras were used in these observations: the Planetary Camera of Wide Field and Planetary Camera 2 (WFPC2/PC), Wide Field Channel of the Advanced Camera for Surveys (ACS/WFC), and the UVIS channel of Wide Field Camera 3 (WFC3/UVIS); their pixel sizes are 0\farcs0455, 0\farcs05, and 0\farcs04, respectively.


\begin{deluxetable*}{ccccccccc}
\tabletypesize{\scriptsize}
\tablecaption{HST Observations}
\tablehead{
PI & Program ID & Filter  & Instrument & Date & Exposure Time\\
 &  &   &  &  (dd/mm/yy) &  (s)}
\startdata 
Morse & 6052 & F547M ($\sim$$V$) & WFPC2/PC & 04/07/1995 & 1000\\
Green & 12001 & F475W ($\sim$$B$) & ACS/WFC & 15/10/2003 & 2280\\ 
Green & 12001 & F550M ($\sim$$V$) & ACS/WFC &  15/10/2003 & 2700\\ 
Green & 12001 & F658N (H$\alpha$) & ACS/WFC &  15/10/2003 & 2160\\
Green & 12001 & F775W ($\sim$$I$) & ACS/WFC & 15/10/2003 & 2160\\
Madore & 12858 & F475W ($\sim$$B$) & ACS/WFC & 10/04/2013 & 2044\\
Milisavljevic & 13378 & F373N ([\ion{O}{2}]) & WFC3/UVIS & 12/05/2014 & 2268\\
Milisavljevic & 13378 & F467M ($\sim$$B$) & WFC3/UVIS & 14/05/2014 & 1362\\
Milisavljevic & 13378 & F502N ([\ion{O}{3}]) & WFC3/UVIS & 13/05/2014 & 2753 
\enddata
\label{table:hst}
\end{deluxetable*}

\subsubsection{Photometric Measurements}

We use the \texttt{DOLPHOT} package to perform point-spread function photometry on the HST images. \texttt{DOLPHOT} is a stellar photometry package adapted from \texttt{HSTPHOT} with HST-specific modules \citep{Dolphin2000}. To select stellar objects, we adopt photometric criteria such as signal-to-noise ratio (S/N) $>$ 5, $sharp^2 < 0.1$, and $crowd < 1.0$, as recommended by \citet{Williamsbf2014}. See \citet{Dolphin2000} for definitions of these parameters. All well-measured stars are cataloged in the Vega magnitude system for the $F475W$ ($\sim$$B$), $F550M$ ($\sim$$V$), and $F775W$ ($\sim$$I$) passbands. We use these photometric measurements to construct color--magnitude diagrams (CMDs) for analyses of the underlying stellar population. 

In the HST continuum-band images, we find that some stars with $V \lesssim$ 18 mag are saturated.  We have thus used the VLT MUSE observations to estimate their magnitudes and colors, as described in the next subsection.

\subsection{VLT MUSE Observations}

The VLT MUSE observations of \E\ were obtained in Program 297.D-5058 (PI: F.P.A. Vogt) on 2016 October 7, using the Director Discretionary Time (DDT) for 9 $\times$ 900 s exposures. The data cube of \E\ has a field of view of 64\farcs6 $\times$ 65\farcs2 with a spatial sampling of 0\farcs2 spaxel$^{-1}$, and a spectral coverage of 4750 -- 9350 \AA\ with a spectral sampling of 1.25 \AA\, pixel$^{-1}$. Details of these observations and procedures of data reduction are described in \citet{Vogt2017a}. 

We use the VLT MUSE data cube to perform spectroscopic analyses and search for stars with peculiar radial velocities as diagnostics of surviving companions. We have also used the VLT MUSE observations in approximate wavelength ranges of $B$, $V$, and $I$ to estimate observed magnitudes and colors of stars that are saturated in HST images. These stars are further added in the CMDs, making the photometric study of stars more complete (see Section 3).

\subsubsection{Extracting Stellar Spectra}

To extract spectra, we use the \texttt{PampelMUSE} package \citep{Kamann2013}, which is designed to optimize the analyses of crowded stellar fields in integral-field spectrograph (IFS) observations.  The stellar sources identified in the high-resolution HST images are used as a mask to guide the MUSE spectral extraction.  \texttt{PampelMUSE} uses point spread function (PSF) fitting to determine the stars' positions and fluxes layer by layer for the data cube. 
The Moffat function, instead of the Gaussian function, is used for the PSF to avoid underestimating the PSF wings.  A fifth order Legendre polynomial is used to smooth the fitted variations across the MUSE wavelength coverage caused by the atmosphere refraction.
Finally, spectra of stars near the central region of \E\ are extracted and saved into individual FITS files. All of the above steps are described in detail in the users' manual of \texttt{PampelMUSE} \citep{Kamann2013}.

As the stars are superposed on the highly nonuniform nebular emission of the SNR,  
it is often difficult to cleanly subtract the nebular background, and the residual
nebular line emission in a stellar spectrum can bias the spectral fits.  Thus, we 
first examine whether residual nebular line emission is still present in the 
sky-subtracted stellar spectra, and exclude the spectral regions that are contaminated 
by nebular lines, such as \oiii $\lambda\lambda$4959, 5007 \AA\ lines. By the same token, 
we have also identify the telluric lines from the observations of the VLT Ultraviolet and 
Visual Echelle Spectrograph (UVES) \citep{Dekker2000}, and exclude them in the spectral fitting. 

\subsubsection{Radial Velocities of Stars and Stellar Parameters}

The radial velocities of stars are determined by cross-correlating stellar spectra with template
spectra generated from the SPECTRUM code \citep{Gray1994}, using the ATLAS9 atmosphere models 
\citep{Castelli2003}.
We select the spectral region 8400 -- 8900 \AA\ that includes hydrogen Paschen
lines and \ion{Ca}{2} triplet for this analysis. In this wavelength range, the 1.25 \AA\ pixel$^{-1}$ spectral sampling of MUSE observations corresponds to $\sim$ 45 \kms\ pixel$^{-1}$. With the signal-to-noise ratios $< 10$ of stellar spectra, a minimum error of 15 \kms\ in radial velocities cannot be ignored \citep{Husser2016, Valenti2018}.
The H$\alpha$ and H$\beta$ lines are avoided because their profiles
are more affected by the background nebular emission.


\subsection{Gaia Observations}
The Gaia Data Release 2 \citep[DR2,][]{Gaia2018} and Early Data Release 3 \citep[EDR3,][]{Gaia2020}
provided precise positions, proper motions, and parallaxes for more than one billion sources 
using the International Celestial Reference System (ICRS).
These data are available in the Gaia Archive (https://archives.esac.esa.int/gaia).
We have used Gaia EDR3 proper motions of stars within \E\
to search for runaway stars as candidates for companion of 
SN progenitor.



\begin{figure*}
\begin{center}
\includegraphics[width=42pc]{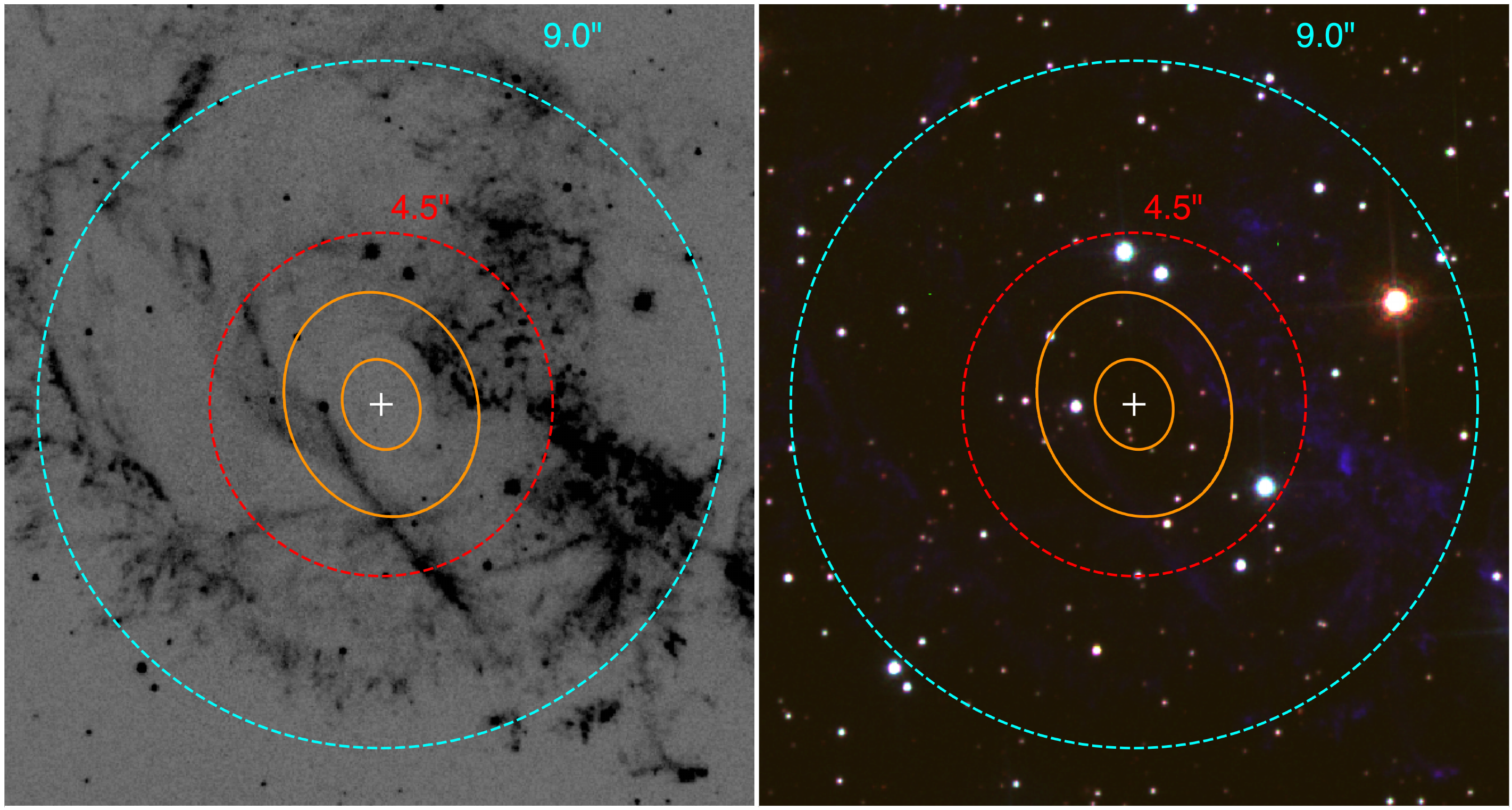}
\caption{ Left: WFC3 F502N image of SNR \E. Right: 
color-composite image of SNR \E, with the F475W image 
in blue, F550M image in green, and F775W image in red.
The two orange ellipses centered at R.A.: 01$^{\mathrm{h}}$04$^{\mathrm{m}}$02$^{\mathrm{s}}$.7; Dec.: $-$72$^\circ$02$'$00\farcs2 (J2000) 
trace the inner and outer edges of the torus of oxygen-rich and neon-rich material \citep{Vogt2018}.
The dashed red and cyan circles illustrate the 4\farcs5 and 9\farcs0 search radii, respectively.}
\label{figure:E0102_runaway_distances2}
\end{center}
\end{figure*}


\begin{figure*}
\begin{center}
\includegraphics[width=42pc]{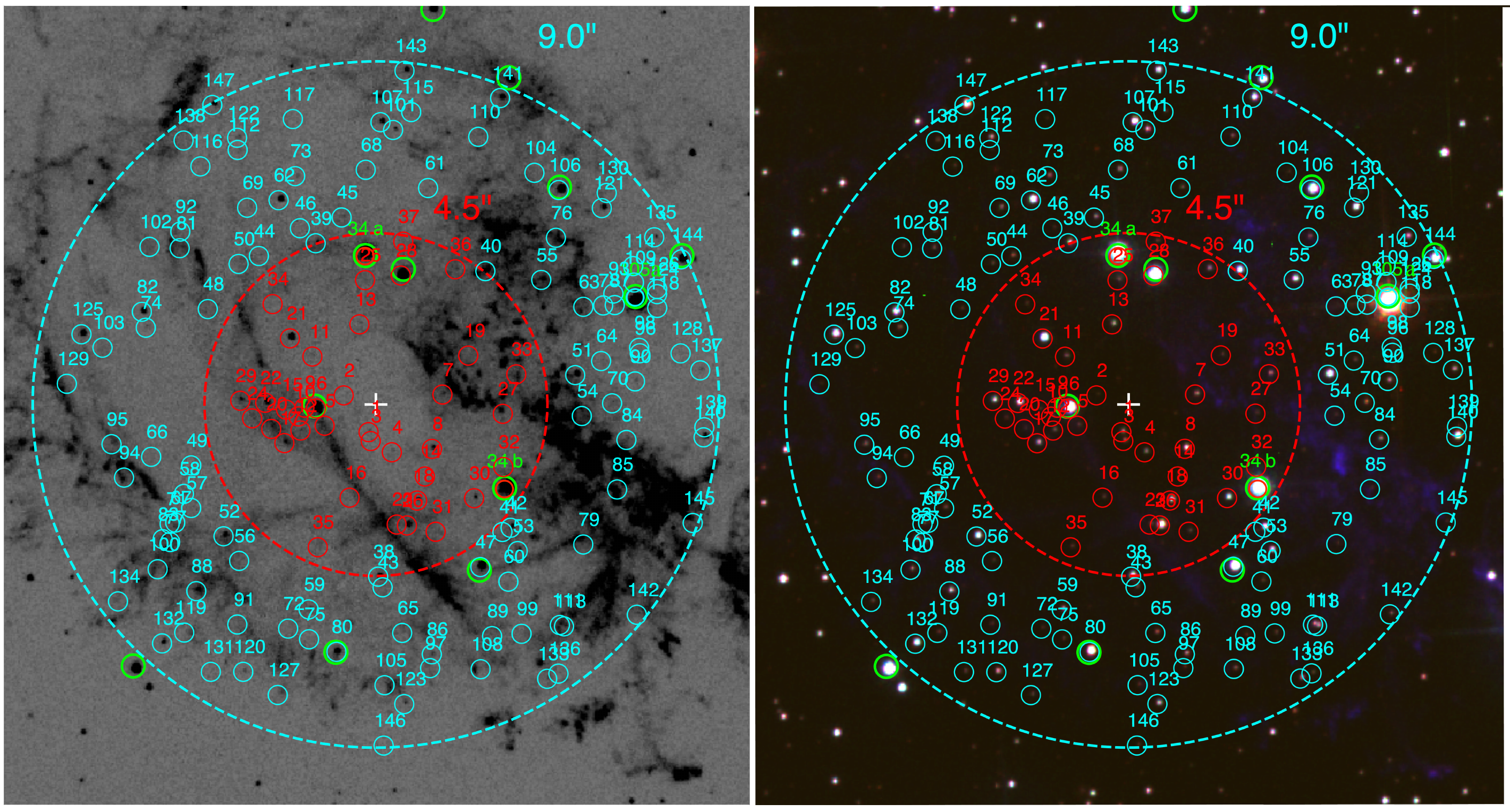}
\caption{Same as Figure \ref{figure:E0102_runaway_distances2}, 
but with stars circled and numbered. 
The stars with $F550M < 26$ within 4\farcs5 and 4\farcs5 to 9\farcs0 from the 
explosion center are circled in red and cyan, respectively.  The stars 
whose proper motions are reported in the Gaia EDR3 are marked with additional green circles.}
\label{figure:E0102_stars}
\end{center}
\end{figure*}

\section{Methodology}

To search for a surviving massive stellar companion of SN E0102's progenitor, we locate the SN explosion site as the center for search, and use the SNR's age multiplied by a generously high runaway velocity as the search radius. All stars within the search radius from the explosion center have been examined to see whether they are viable candidates for a surviving companion of \E's SN progenitor. We use properties of companions of Type IIb SN progenitors reported in the literature to guide our search.  Finally, we examine the proper motions and radial velocities of stars to diagnose runaway stars as candidates of SN progenitor's companion.

\begin{figure*}[htbp]
\subfigure[]{
\hspace{-0.5cm}
\includegraphics[scale=0.6]{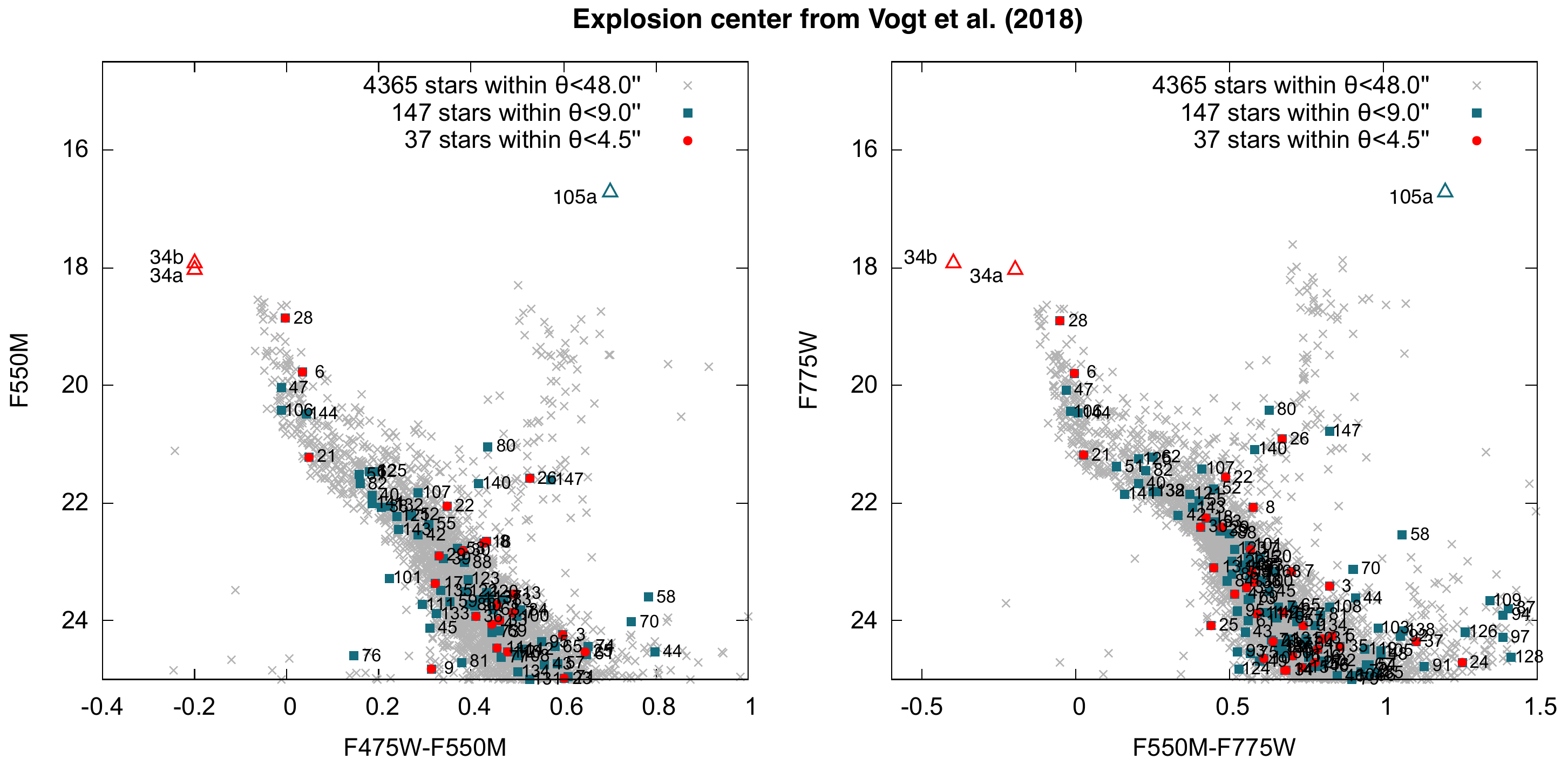}
}\\%
\subfigure[]{
\hspace{-0.5cm}
\includegraphics[scale=0.6]{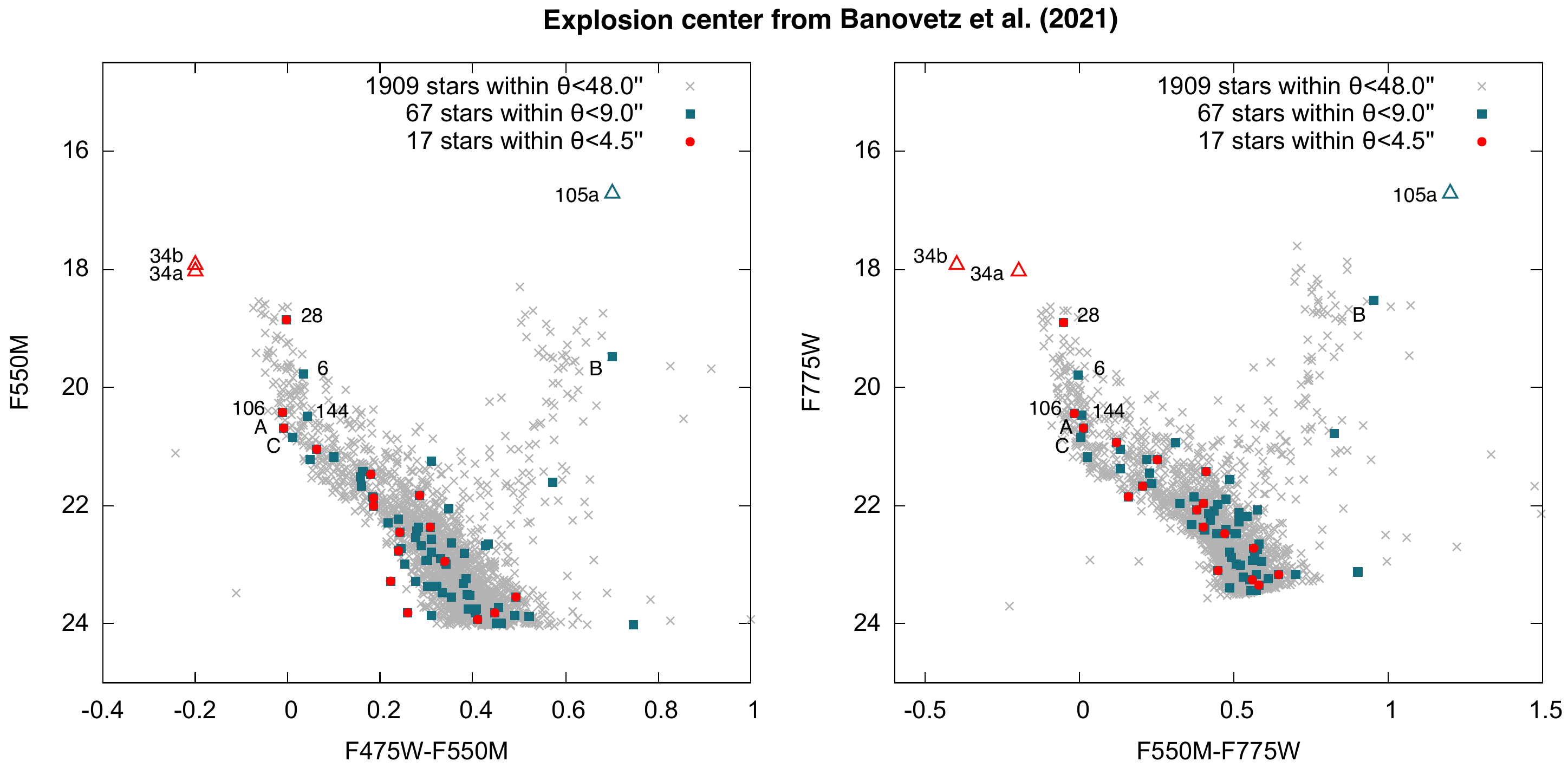}
}%
\caption{
Color-magnitude diagrams of stars in E0102 constructed 
from HST photometric data.  The left panels are 
plots of F550M versus F475W $-$ F550M and the right panels F775W versus F550M $-$ F775W.
Stars within 4\farcs5 and 9\farcs0 of the explosion center are 
plotted in red filled circles and blue filled squares, respectively. 
To illustrate the main sequence and red giant branch of the underlying
stellar population, all stars within 48\farcs0 from the SN explosion 
center in \E\ are plotted in grey crosses.
Photometric measurements of bright stars that are saturated in HST images are made from 
the VLT MUSE data cube and plotted in open triangles, following the same color scheme,
red for stars within 4\farcs5 of the explosion center and blue for 4\farcs5 to 9\farcs0. 
(a) uses the SN explosion site suggested by \citet{Vogt2018} and (b) uses the SN explosion 
center reported by \citet{Banovetz2021}. A brighter cutoff of F550M $<$ 24 is used in (b) 
to avoid cluttering.}


\label{figure:E0102_CMDs_HST_muse}
\end{figure*}

\subsection{Possible Binary Companion and Search Area}

To assess the SN explosion site, \citet{Finkelstein2006} used the Chandra \xray images to determine a geometric center and used the proper motion of SN ejecta to derive an expansion center of \E. Recently, \citet{Vogt2018} discovered a small slowly expanding optical ring of oxygen-rich and neon-rich material, as marked in Figure \ref{figure:E0102_runaway_distances2}, and a point X-ray source near the center of this ring.  These results led the authors to suggest the center of the optical ring at 01$^{\mathrm{h}}$04$^{\mathrm{m}}$02$^{\mathrm{s}}$.7, $-$72$^\circ$02$'$00\farcs2 (J2000) to be the SN explosion site, also marked in Figure \ref{figure:E0102_runaway_distances2}.  This center is about 5\farcs9 (1.77 pc) southeast of the expansion center suggested by \citet{Finkelstein2006}.  \citet{Vogt2018} argue that it is unlikely for the NS and the small ring of metal-rich material to have run away from Finkelstein et al.'s explosion center at a transverse velocity of $\sim$850 \kms.
Even more recently, \citet{Banovetz2021} analyzed the proper motions of the oxygen-rich SN ejecta from HST images taken in 4 epochs over a time span of 19 yr, and determined an explosion center that is closer to that of \citet{Finkelstein2006}.  The explosion centers suggested by the above three groups are marked in Figure~\ref{figure:E0102_F502N} by crosses and labeled by F, V, and B, respectively.
In this paper, our analysis is made first with the explosion center suggested by \citet{Vogt2018}, then repeated for the explosion center suggested by \citet{Banovetz2021}.

After the SN explosion, the stellar companion can become unbound and run away from the SN explosion site. In general, the runaway velocity of an unbound companion can be from tens \kms\ \citep{Tetzlaff2011, Dincel2015, Dufton2011, Renzo2019} to hundreds of \kms\ \citep{Hoogerwerf2001, Eldridge2011}, with the greatest majority well below 300 \kms.
In a rare case, the B star HD 271791 has been observed to have a velocity of 530--920 \kms\ in the Galactic-rest frame \citep{Heber2008} and the observed enhancement of $\alpha$-elements in its atmosphere suggests that it is a SN's surviving companion \citep{Przybilla2008}.  Theoretically, it may be possible for companions to reach runaway velocities $\geq$ 1,000 \kms\ \citep{Renzo2019}, but it is only when the geometry and the energy of SN explosion and the mass of the companion are under the most extreme favorable conditions \citep{Tauris2015}. 

The ``runaway distance'' of a surviving companion from the SN explosion site is equal to the runaway velocity times the SNR's age. We adopt the age of 1735$\pm$175 yr determined by \citep{Banovetz2021}, and conservatively use search radii of 4\farcs5 (\s1.35 pc) and 9\farcs0 (\s2.7 pc), which correspond to maximum runaway velocities of 760 and 1500 km s$^{-1}$, respectively. 
These two search radii are illustrated in Figure \ref{figure:E0102_runaway_distances2}, which assumes the explosion center suggested by \citet{Vogt2018}.

\subsection{Consideration of Possible Types of Companion}

Massive binary systems usually have secondary-to-primary mass ratios greater than 0.5 
\citep{Pinsonneault2006, Sana2012}; thus, the secondary that survives a CCSN explosion
is likely a massive star.  Furthermore, because of the mass loss of the SN ejecta the
binary system is likely to become unbound and the surviving companion (the secondary) would 
become a runaway OB star.  Among the CCSNe, \iib SNe belong to the subset whose 
progenitors' envelopes have been stripped prior to the SN explosion. 
For a subsolar metallicity similar to that of the SMC, binary population synthesis 
simulations show that among all stripped-envelope CCSNe, 2/3 have main-sequence companions
to their progenitors, about 1/4 have single progenitors at death, and only less than 5\% 
progenitors have giant or compact companions \citep{Zapartas2017b}. 

Observationally, the companions of \iib SNe have been shown to be hot and luminous companions 
from pre- and post-explosion images, for example, SN 1993J \citep{Maund2004, Stancliffe2009, Fox2014}, SN 2008ax \citep{Folatelli2015}, SN 2011dh \citep{Bersten2012, Benvenuto2013, Folatelli2014}, and SN 2001ig \citep{Ryder2018}, although some evidence may be weak \citep{Maund2015}. The above studies indicate that the zero-age main sequence (ZAMS) masses of these SNe's progenitors are around $\sim$ 10--22 \msun\, while the final masses of their companions before the SN 
explosion are in the range of $\sim$ 10--25 \msun\ and their temperatures are roughly 16,000--40,000 K and 
luminosities higher than 6,000 \lsun. 

To simulate SN 1993J's progenitor,  \citet{Claeys2011} adopted a ZAMS mass of 15 \msun\ for the primary to compute an extensive grid of binary models,
and suggested three possible types of companion star for \iib SNe at the time of SN explosion: hot O star in 
$\sim$90\% of the cases, over-luminous B star in $\sim$3\% of the cases, and K supergiant in $\sim$7\%
of the cases where the primary and secondary have very similar initial masses. They have also provided the expected broad-band photometry for these three types of companions, e.g., $M_V$ $\sim$ $-$4.65 mag, $-$6.22 mag, and $-$6.14 mag, respectively. Similarly, \citet{Sravan2020} used single and binary models to probe a wider parameter space for \iib SNe. and found a dominant fraction of blue MS companions with 500--160,000 \lsun\ and a smaller fraction of red and yellow evolved companions with 25,000--250,000 \lsun. 

The results of above studies of \iib SN progenitors can guide our search for a surviving companion in \E.  
The impact of SN ejecta can shock-heat and even strip some of the companion's atmosphere, so
the post-impact evolution of the companion needs to be considered. Fortunately, in most cases, the CCSN explosion is expected to have no dramatic effects on the post-impact stellar evolution of the companion \citep{Stancliffe2009, Liu2015, Rimoldi2016, Liu2019}. Using hydrodynamical simulations, \citet{Hirai2018} show that the surviving companions of CCSNe can be an order of magnitude overluminous right after the SN explosion, but this high luminosity fades away after $\sim$ 10 yr and would not last forever. 

In summary, we search for a surviving blue companion with $>$ 500 \lsun\ or a surviving yellow or red evolved companion with $>$ 25,000 \lsun\ within E0102, corresponding to 
OB stars and K supergiants, respectively. We plot both stars in our 4\farcs5 and 9\farcs0 search radii and stars within a large radius encompassing the entire SNR in the CMDs. The stars in the largest radius are used to illustrate the underlying stellar population. We then compare magnitudes and colors of the stars in our search radii with those expected from surviving companions of Type IIb SNe. The stars having photometric properties that are consistent with our expectations are candidates of surviving companions.

\subsection{Stellar motions}

Most binary companions of SN progenitors would become unbound after the SN explosions \citep{Renzo2019},
resulting in runaway stars.
A runaway star can be diagnosed by its 
peculiar proper motion or radial velocity.  Thus, we use the HST images 
and Gaia EDR3 to investigate proper motions of stars projected in and
near \E. We also use VLT MUSE observations to assess radial velocities of 
stars near the SN explosion site.

\subsubsection{Proper Motions of Stars}

To assess stellar proper motions from HST images, we first use the \texttt{DOLPHOT} package to measure the positions of stars in each image. We then adopt a method similar to that used by \citet{Kerzendorf2019} to align the stars in images from different epochs in order to search
for stars with large proper motions as candidates of runaway companion of the SN progenitor.
Among the archival HST imaging observations of \E, the longest time lapse between epochs is $\lesssim$ 19 yr. Unfortunately, the expected proper motion of a companion is too small for the HST data to be useful, as demonstrated later in Section 4.2.1. The HST observations we used are listed in Table \ref{table:hst}. 

The Gaia EDR3 provides parallax and proper motion information for stars.  The parallax
of the SMC is small and the errors are large, so we only use the parallax information to remove 
foreground stars in the Milky Way.  We use the 62 kpc distance to the SMC to convert the proper motion reported in Gaia EDR3 to tangential velocities of stars within \E. The stars that have large tangential velocities and are located in the SMC would be candidates of surviving companions. 

\subsubsection{Radial velocities of Stars}

We have used VLT MUSE observations to perform spectroscopic analyses of stars within our search radii.  The observed spectra are fitted by stellar atmospheric models to determine the stellar parameters, such as effective temperature, luminosity, and extinction.  We use Gaussian fittings of line profiles and cross-correlations between model spectra and observed spectra to determine the radial velocities of stars with $V$ $<$ 21.5 mag. This upper limit of $V$ corresponds to a $\sim$2 $M_\odot$ star in the SMC.  It is a fairly safe limit for our search for surviving companions of \iib SNe, which are expected to be brighter than 500\lsun\ in most cases \citep{Maund2004, Claeys2011, Sravan2020}. We thus examine the spectra of photometrically selected candidates of surviving companions (i.e., those with $V$ $<$ 21.5 mag and near the SN explosion site) and use large radial velocities to further evaluate their nature and advance their candidacy as surviving companion of the SN progenitor.

\begin{table*}
\caption{Stellar parameters of the stars with $V < 21.5$ mag within 9\farcs0 from Vogt et al.\ explosion center}
\centering
\begin{tabular}{@{}ccccccccccccccc@{}}
\hline\hline
Star & $V_{\textrm{r}}$ & $T_{\textrm{eff}}$ & log$_{10}\ L$ & log$_{10}\ g $ & $V$ & $B$--$V$ & $V$--$I$ & Spectral \\
  & (\kms) & (K) & (L$_\odot$) & (dex)  & (Spectral) & (Spectral) & (Spectral) & Type \\
\hline
 6 & 140 $\pm$ 35 & 12250 $\pm$ 900 & 1.83 & 4.5 & - & - & - & B8V &  \\
 21 & 140 $\pm$ 65 & 10250 $\pm$ 760 & 1.18 & 4.5 & - & - & - & B9.5V &  \\
 28 & 155 $\pm$ 35 & 14000 $\pm$ 1210 & 2.27 & 4.5 & - & - & - & B7V &  \\
 34a & 175 $\pm$ 30 & 17000 $\pm$ 1715 & 2.79 & 4.0 & 18.0 & -0.2 & -0.2 & B3V &  \\
 34b & 190 $\pm$ 40 & 24000 $\pm$ 3050 & 3.21 & 4.5 & 17.9 & -0.2 & -0.4 & B1.5V &  \\
\hline
 47 & 215 $\pm$ 30 & 16000 $\pm$ 1900 & 1.91 & 4.5 & - & - & - & B5V &  \\
 62 & 150 $\pm$ 50 & 8000 $\pm$ 335 & 0.96 & 4.5 & - & - & - & A6V &  \\
 80 & 165 $\pm$ 20 & 6000 $\pm$ 120 & 1.11 &  4.5  & - & - & - & F9.5V &  \\ 
 105a & 158 $\pm$ 15 & 4800 $\pm$ 120 & 2.85 & 2.0 
& 16.9 & 0.6 & 1.3 & K0II &  \\
 106 & 170 $\pm$ 35 & 12750 $\pm$ 1110 & 1.65 & 4.5 & - & - & - & B8V &  \\
 125 & 150 $\pm$ 20 & 7750 $\pm$ 265 & 0.93 & 4.5 & - & - & - & A7V &  \\
 144 & 180 $\pm$ 35 & 14000 $\pm$ 1300 & 1.66 & 4.5 & - & - & - & B7V &  \\
\hline
\multicolumn{8}{l}{Note: stars with number $<$ 38 are within the 4\farcs5 search radius from the SN explosion site.}
\end{tabular}
\label{table:stellar_param}
\end{table*}

\begin{table*}
\caption{ Stars within 9\farcs0 from Vogt et al.'s SN Explosion Site Detected in Gaia EDR3}
\centering
\begin{tabular}{@{}ccccccccccccc@{}}
\hline\hline
Star & R.A.  & Decl. & Parallax & $\mu_\alpha$ & $\mu_\delta$  & $\mu_\alpha-\mu_\alpha^{\mathrm{SMC}}$ & $\mu_\delta-\mu_\delta^{\mathrm{SMC}}$ & r$_V$  & r$_B$\\
 & (J2000) &  (J2000) & (mas) & (mas yr$^{-1}$) & (mas yr$^{-1}$) & (mas yr$^{-1}$) & (mas yr$^{-1}$) &  (arcsec)  &  (arcsec)\\
\hline
6 & 1:04:03.04 & -72:02:00.27  & -0.383  $\pm$  0.416  &  1.22  $\pm$  0.47  &  -1.33  $\pm$  0.46  &  0.41  $\pm$  0.50  &  -0.11  $\pm$  0.50 & 1.58 & 6.86\\
28 & 1:04:02.55 & -72:01:56.78  & \, 0.540  $\pm$  0.213  &  1.29  $\pm$  0.27  &  -1.63  $\pm$  0.33  &  0.48  $\pm$  0.33  &  -0.41  $\pm$  0.38 & 3.49 & 2.88\\
\,\,  34a & 1:04:02.76 & -72:01:56.21  & -0.157  $\pm$  0.111  &  0.64  $\pm$  0.13  &  -1.04  $\pm$  0.13  &  -0.17  $\pm$  0.23  &  0.18  $\pm$  0.22 & 4.00 & 2.63 \\
\,\, 34b & 1:04:01.96 & -72:02:02.39  & 0.062  $\pm$  0.094  &  0.78  $\pm$  0.12  &  -1.28  $\pm$  0.11  &  -0.03  $\pm$  0.22  &  -0.06  $\pm$  0.21 & 4.06 & 8.81 \\
\hline
47 & 1:04:02.10 & -72:02:04.45  & -0.440  $\pm$  0.486  &  1.46  $\pm$  0.64  &  -0.59  $\pm$  0.61  &  0.65  $\pm$  0.66  &  0.63  $\pm$  0.63 & 5.08 & 10.68\\
80 & 1:04:02.92 & -72:02:06.70 & -   & -  &  - & - & -  & 6.58 & 12.94\\
\,\, 105a & 1:04:01.22 & -72:01:57.52  & -0.050  $\pm$  0.047  &  0.83  $\pm$  0.06  &  -1.22  $\pm$  0.06  &  0.02  $\pm$  0.20  &  0.00  $\pm$  0.19 & 7.35 & 6.85\\
106  & 1:04:01.65 & -72:01:54.53 & -0.917  $\pm$  0.613  &  1.00  $\pm$  0.91  &  -0.98  $\pm$  1.07  &  0.19  $\pm$  0.93  &  0.24  $\pm$  1.08 & 7.47 & 3.89\\
144 & 1:04:00.96 & -72:01:56.36  & 2.930  $\pm$  0.981  &  3.43  $\pm$  1.26  &  0.48  $\pm$  1.30  &  2.62  $\pm$  1.27  &  1.70  $\pm$  1.31 & 8.92 & 7.45\\
\hline
A & 1:04:02.38 & -72:01:49.78 & -0.086  $\pm$  1.013  &  0.39  $\pm$  1.60  &  -0.57  $\pm$  1.45  &  -0.42  $\pm$  1.61  &  0.65  $\pm$  1.46 & 10.53 & 4.17 \\
B & 1:04:01.97 & -72:01:48.59 & 0.036  $\pm$  0.235  &  0.86  $\pm$  0.28  &  -0.91  $\pm$  0.29  &  0.05  $\pm$  0.34  &  0.31  $\pm$  0.34 & 12.09 &  5.83 \\
C & 1:04:03.67 & -72:01:47.10 & -0.171  $\pm$  1.206  &  4.15  $\pm$  1.71  &  -1.25  $\pm$  2.00  &  3.34  $\pm$  1.72  &  -0.03  $\pm$  2.01 & 13.85 & 8.77  \\
\hline
\multicolumn{10}{l}{Note: The running numbers of stars are from Table~\ref{table:photometryGreen}.  r$_V$  and r$_B$ are distances to the SN explosion site suggested by}\\
\multicolumn{10}{l}{\citet{Vogt2018} and \citet{Banovetz2021}, respectively. The bottom three stars are in Table~\ref{table:photometryNew} but not in Table~\ref{table:photometryGreen}.}
\end{tabular}
\label{table:gaia}
\end{table*}

\section{Results}

To search for a surviving companion within SNR \E, we have used photometric measurements of stars projected within and near this remnant to construct CMDs, and compare locations of the stars near the SN explosion site in the CMDs with expected surviving companion. We have also analyzed stellar motions of our candidates including their proper motions and radial velocities. Below we report our findings of these two methods for \citet{Vogt2018} explosion center in Section 4.1 and
for \citet{Banovetz2021} explosion center in Section 4.2. 

\subsection{The Case for Vogt et al.\ Explosion Center }

\subsubsection{Candidates Selected from Color-Magnitude Diagrams}

We have marked a total of 150 stars with $V < 26$ mag within 9\farcs0 search radius in the close-up \ha and color-composite images in Figure \ref{figure:E0102_stars}, with stars within 4\farcs5 in red circles and stars from 4\farcs5 to 9\farcs0 in cyan circles. These 150 stars are numbered in the order of increasing angular distance from \citet{Vogt2018} SN explosion site. Among them, 147 stars are well-measured in the HST continuum-band images and their magnitudes and colors are given in the Appendix \ref{appendix:photometry}. The other three stars, 34a, 34b, and 105a, are saturated in the HST continuum-band images, thus photometric measurements from the VLT MUSE observations (see Section 2) are adopted, as listed in Table \ref{table:stellar_param}. 

We have plotted these 150 stars in the equivalents of $V$ versus $B-V$ ($F550M$ versus $F475W-F550M$) and $I$ versus $V-I$ ($F775W$ versus $F550M-F775W$) CMDs in Figure~\ref{figure:E0102_CMDs_HST_muse}a with stars within 4\farcs5 in red symbols and
stars from 4\farcs5 to 9\farcs0 in blue symbols, following the same color scheme as 
Figure~\ref{figure:E0102_stars}.  The locations of these stars in the CMDs are compared with those of expected companions of \iib SNe's progenitors, for which we use a distance modulus of 18.95 $\pm$ 0.07 (corresponding to 62 kpc) for the SMC \citep{Graczyk2014} and a color excess $E(B-V)$ $=$ 0.08 toward \E\ \citep{Blair1989}. 

Within 4\farcs5 from the SN explosion site, a total of 39 stars are brighter than $V < 26$ mag. We first examine MS stars for possible companion candidates. As the mass distribution of MS companions of stripped-envelope SN progenitors peak around 9 \msun\ \citep{Zapartas2017b}, we consider the five most massive MS stars: 34a, 34b, 28, 6, and 21.  
Stars 34a and 34b are the hottest and most luminous, and their $B-V$, $\sim$\ $-$0.2 and $-$0.1, both suggest early B spectral types. Their $M_V$ magnitudes, $\sim$\ $-$1.05 mag and $-$1.15 mag, are consistent with B type stars in the SMC. Compared with the over-luminous B companion of $M_V$ = $-$6.22 mag at the time of SN explosion \citep{Claeys2011}, stars 34a and 34b are $\sim$110 and $\sim$100 times fainter and thus not over-luminous. Although the masses and the luminosities of stars 34a and 34b are close to the expected range of \citet{Zapartas2017b} and \citet{Sravan2020}, these two stars are ultimately rejected due to their proper motions, as discussed in Section 4.1.2. 
Stars 28, 6, and 21 have $B-V$ colors, $\sim$ $-$0.08, $-$0.05, and $-$0.03, respectively, indicating late-B spectral types that are consistent with their $M_V$.  These late B stars have masses of a few $M_\odot$ and are unlikely to be companions.  Beside these five B stars, the remaining MS stars within 4\farcs5 from the SN explosion site have spectral types roughly from A to K with masses $\lesssim$ 3 \msun\
(Figure \ref{figure:E0102_CMDs_HST_muse}), and they are even less likely to be candidates for MS companions of a stripped-envelope SN \citep{Zapartas2017b}. 

Within 4\farcs5 from the explosion center, star 26 is the only star on the red giant branch (RGB), as shown in Figure \ref{figure:E0102_CMDs_HST_muse}. This star's $B-V$ $\sim$ 0.45 is suggestive of a F5-F6 spectral type, and its location in the CMD indicates that it is a mid-F subgiant evolved from stars with initial masses of $\sim$2 \msun. If this is the SN progenitor's companion, the primary to secondary mass ratio would exceed 4, which is not likely.
Furthermore, \citet{Zapartas2017b} suggest that giant companions only contribute to a very small portion (\s0.5\%) of stripped-envelope SNe, as the parameters of initial binary system need to be fine-tuned. Thus, we do not consider star 26 a plausible candidate for surviving companion. 

Within 4\farcs5 to 9\farcs0 from the explosion site, stars 47, 106, and 144 are the only three blue stars on the MS, with $B-V$, $\sim$ $-$0.01, $-$0.01, and 0.04, respectively (see Figure \ref{figure:E0102_CMDs_HST_muse}).  They are of mid- to late-B spectral types. Their masses are more or less consistent with the MS companions of stripped-envelope SNe \citep{Zapartas2017b}; however, their distances to the SN explosion site, $\sim$ 5\farcs1, 7\farcs4, and 8\farcs9, indicate a large runaway velocity $\sim$ 860 \kms, 1250 \kms, and 1500 \kms, respectively. While such high velocities cannot be ruled out \citep{Tauris2015, Heber2008, Przybilla2008}, they are uncommon for CCSN companions and highly unlikely \citep{Hoogerwerf2001, Eldridge2011, Renzo2019}. Outside the MS, three stars are observed to be in the mid-F subgiant region in the CMD, and we do not consider these low-mass evolved stars to be viable candidates for SN progenitor companion. Last but not least, we notice that star 105a is located above the RGB and is the most luminous star in the CMDs. Its locations in the CMDs are not well populated by SMC or Galactic stars (Figure \ref{figure:E0102_CMDs_HST_muse}). We will further discuss this star with its spectral analyses consideration in Section 4.1.3.


\begin{figure}[h]  
\hspace{-0.5cm}
\includegraphics[width=23pc]{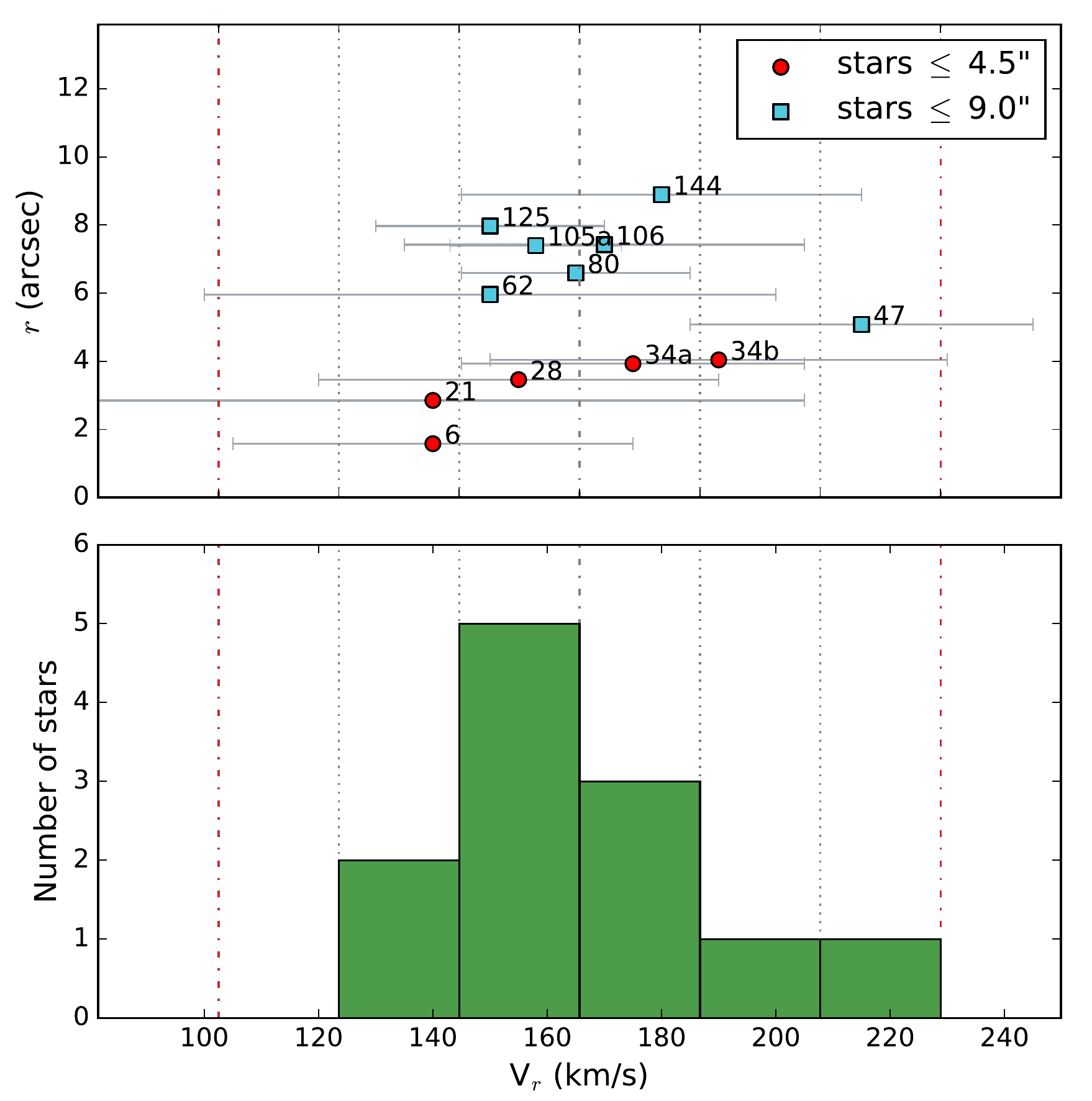}
\caption{
Top: plot of radial velocity (Vr) with uncertainties versus the angular distance to the SN explosion site (r) for stars with V $<$ 21.5 mag within 9\farcs0 in \E\ (see Table \ref{table:stellar_param}). Bottom: cumulative number of these stars within standard deviations ($\sigma$) from the mean radial velocity (Vr).
}
\label{figure:E0102_vrad.pdf}
\end{figure}

\subsubsection{Stellar Motions}

We first consider proper motions of stars using the large number of HST images of \E\ in the archive.  A partial list is given in Table~\ref{table:hst}.  The 1995 and 2014 images form the longest baseline in time for proper motion measurements.  While proper motions for SN ejecta have been successfully measured and reported by \citet{Finkelstein2006} and \citet{Banovetz2021}, the proper motion of a runaway star is much harder to measure because the expansion velocity of the SN ejecta is more than 20 times faster than the expected velocity of a runaway star, well below 300 \kms\ \citep{Eldridge2011}. A transverse velocity of 300 km s$^{-1}$ at the SMC distance of $\sim$ 62 kpc corresponds to a proper motion of $\sim$1 mas yr$^{-1}$; from 1995 to 2014, this transverse velocity produces an angular displacement of $\sim$ 19 mas.  In \texttt{DOLPHOT}, the astrometric uncertainties are typically $\sim$ 10 to 60 mas for bright to faint stars, respectively \citep{Dolphin2000}. Therefore, it would be challenging to use the currently available HST observations to carry out a reliable search of large proper motions of stars.


We then resort to Gaia EDR3 for proper motions of stars. Within 9\farcs0 from \citet{Vogt2018} explosion center a total of nine stars with $V$ $\lesssim$ 21.0 mag are reported in the Gaia EDR3, as marked in Figure \ref{figure:E0102_stars}. To search for peculiar motion of these stars, the proper motion of the SMC needs to be corrected from the apparent proper motion given by Gaia EDR3.  We have found Gaia proper motion for 398 stars within 1$'$ from \E.  After removing stars with proper motion errors greater than 100 km s$^{-1}$ and stars with proper motion deviating from the median by 3$\sigma$, only 113 stars remain and are used to determine the local mean proper motion, $\mu_\alpha$ = $0.81 \pm 0.19 $ mas yr$^{-1}$ and $\mu_\delta$ = $-1.22 \pm 0.18$ mas yr$^{-1}$, consistent with the values $\mu_\alpha$ = $0.874 \pm 0.066 $ mas yr$^{-1}$ and $\mu_\delta$ = $-1.2047 \pm 0.33$ mas yr$^{-1}$ reported by \citet{vanderMarel2016} using Gaia DR1. 
In Table \ref{table:gaia}, we list the nine stars' positions, Gaia parallaxes and proper motion, 
proper motion relative to the local field, and distances to the explosion site (arcsec). Apparently only star 144 has a $>$ 2$\sigma$ detection of proper motion.  The Gaia parallaxes are clearly not reliable, so we use the 62 kpc distance for the stars.  Then the proper motion of star 144 is more than 770 km s$^{-1}$ to the east.  Such a high velocity is either spurious or extremely interesting.  Considering that the radial velocity of star 144, $180 \pm 35$ km s$^{-1}$, is offset from the SMC velocity by less than 20 km s$^{-1}$, it is perhaps more likely a spurious measurement.



We have analyzed the spectra of all 12 stars with $V<$21.5 mag within 9\farcs0 from the \citet{Vogt2018} explosion center to search for peculiar radial velocities as diagnostics of surviving companions. The radial velocities and physical parameters of these 12 stars from our fittings are listed in Table \ref{table:stellar_param}. 

We have plotted the radial velocities of these 12 stars versus their distance to the SN explosion site in Figure \ref{figure:E0102_vrad.pdf}.  The mean radial velocity is $166 \pm 21$ km s$^{-1}$, consistent with the bulk velocity of the SMC.  The error bar ($\sigma$) includes both observational errors and dispersion among the stars. In Figure \ref{figure:E0102_vrad.pdf} we have over-plotted vertical lines at the mean and 1--3$\sigma$ from the mean, and it can be seen that no stars deviate from the mean by $\ge$ 3$\sigma$. Star 47, with a radial velocity of 215$\pm$30 \kms, has the largest deviation from the mean; however, its distance to the SN explosion site, $\sim$5\farcs1, requires a runaway velocity of $\sim$860 \kms\ that is too high to make it a plausible candidate for surviving companion.


\subsubsection{The Case of Star 105a}

In the optical bands, star 105a is the brightest star within the SNR \E\ and its surroundings within 48\farcs0 from \citet{Vogt2018} SN explosion site (Table~\ref{table:photometryGreen} in Appendix \ref{appendix:photometry}).
Its location in the CMDs (Figure~\ref{figure:E0102_CMDs_HST_muse}) is above the RGB in a region not well populated by SMC or foreground stars. Its absolute magnitudes and colors suggest a spectral type of K0\,II, consistent with the strengths of the \ion{Ca}{2} triplet in its spectrum and a stellar effective temperature of $\sim$4800 K determined from atmospheric model fits to its spectrum.  See Table~\ref{table:stellar_param} for the model-fitted stellar parameters.  

As \citet{Claeys2011} predicted the existence of surviving K supergiants, we compare the K0\,II star 105a to a K supergiant.  A K0 supergiant has $M_V$ $\sim$ $-$6.1 mag, while star 105a has $M_V$ $\sim$ $-$2.25 mag, about 35 times fainter than a K0 supergiant.
Comparisons of the temperature and luminosity of star 105a with stellar evolutionary tracks indicate that it is an evolved star with an initial mass of 4-5 $M_\odot$. Even if 105a has accreted more than 5\% of its mass from the primary companion and spun up, the rejuvenation caused by rotational mixing implies that it is actually older than what a single 4-5 $M_\odot$ star \citep{Eldridge2017}, the MS lifetime of 105a is more than 90 Myr, much longer than the lifetime of the SN progenitor.  Further, neither the radial velocity (Figure \ref{figure:E0102_vrad.pdf}) nor the proper motion (Table \ref{table:gaia}) of star 105a exhibits large anomaly.
Thus, we do not consider star 105a a likely companion of the SN progenitor of \E.

\subsection{The Case for Banovetz et al.\ Explosion Center}

\subsubsection{Candidates Selected from Color-Magnitude Diagrams}

We have compiled all stars with $V <$ 24 within 9$''$ of \citet{Banovetz2021} explosion center in Table~\ref{table:photometryNew}.  To avoid confusion, we do not provide new running numbers; instead, running numbers from 
Table~\ref{table:photometryGreen} are used for the few stars that are discussed here.  
These stars are used to construct CMDs in Figure~\ref{figure:E0102_CMDs_HST_muse}b, where stars within 4\farcs5 of the explosin center are plotted in red filled circles, and stars within 4\farcs5 to 9$''$ of the explosion center in blue filled squares.  For stars that are saturated in HST images, we use the photometry extracted from the MUSE data and plot them in open triangles according to the same color scheme used for the stars with HST photometry. 

Following the rationale of companion search used in Section 4.1.1, we first consider the brightest MS stars, with $M_V < 20$,
because fainter stars, such as 106 and 144, are of late B-types with masses $< 5 M_\odot$ and are unlikely to be a companion of the SN progenitor.  The MS stars that warrant discussion are 34a, 34b, 28, and 6.  For Banovetz et al.'s SN explosion site, star 34a is the nearest; its distance of 2\farcs63 (0.79 pc) to the explosion center requires a transverse velocity of $445 \pm 45$ km s$^{-1}$ for a SN age of 1738 $\pm$ 175 yr \citep{Banovetz2021}.  While the transverse velocity of 34a is on the high side, its spectral type, B3\,V, implies a mass of 7.6 $M_\odot$.  These properties make 34a a possible candidate and will be discussed in more detail in Section 4.2.3.   Among the other three stars, stars 6 and 34b are too far away from the explosion site and require transverse velocities $\gtrsim$1000 km s$^{-1}$, and star 28 is a B7\,V star with a 4.4 $M_\odot$ mass that is too low to be a candidate. 

\subsubsection{Stellar Motions}

Gaia EDR3 proper motion data for Stars within 9$''$ from Banovetz et al.'s SN explosion site are listed in Table~\ref{table:gaia}.  The bottom three stars are in Table~\ref{table:photometryNew} but not in Table~\ref{table:photometryGreen}; we designate these stars as A, B, and C and mark them in Figure~\ref{figure:E0102_CMDs_HST_muse}b. Among the stars that have Gaia proper motion data, we use the distance to the SN explosion center divided by the SNR's age to determine the expected proper motion, and exclude the ones whose Gaia proper motions are far lower than the expected.  Only four stars remain: stars 28, 34a, 144, and C.  While stars 144 and C cannot be immediately rule out, their expected transverse velocities are $>$1000 km s$^{-1}$, and hence very unlikely.  Only stars 28 and 34a cannot be immediately dismissed from the comparisons between their expected transverse velocities and Gaia measured proper motions.  Star 28 is a B7\,V star, $\sim$4.5 $M_\odot$, and 34a is a B3\,V star of $\sim$7.6 $M_\odot$.

As shown in Table~\ref{table:stellar_param}, only star 47 has radial velocity significantly higher than the other stars; however, star 47 is too far from the explosion center to be considered as a candidate.  The radial velocities, being similar among all stars, do not support or reject any star as a surviving companion of the SN.

\subsubsection{The Case of Star 34a}

Star 34a is the nearest massive star to Banovetz et al.'s SN explosion site.  Based on the mass and kinematics considerations, star 34a not only cannot be excluded, but also has the most optimal properties matching the expectations of a surviving companion.  It is interesting to note a peculiar feature in star 34a's spectrum, shown in Figure~\ref{figure:Star34a}. The spectral region near 5000 \AA\ is contaminated by the [\ion{O}{3}] $\lambda\lambda$4959, 5007 lines and the imperfect background subtraction results in artificial emission/absorption spikes.  However, there is a broad bump in the continuum extending from $\sim$5000 to $\sim$5300 \AA. 
While this feature is not seen in the spectrum of star 34b in Figure~\ref{figure:Star34b}, it is also seen in the spectrum of the early B variable star Cl* NGC\,371 KAG V28 \citep{Karoff2008} southwest of E0102.  It is not clear whether this broad bump is an artifact or real. Follow-up spectroscopic observations are needed to confirm this spectral feature and to assess its abundance, radial velocity, and rotational velocity more accurately in order to strengthen or reject its nature as the surviving companion of the SN progenitor of E0102.



\section{Discussions}


We will discuss the OB star binary companion of E0102's SN progenitor and origins of the NS.

\subsection{OB Star Companion}

Type IIb SNe require their progenitors' hydrogen envelopes to be stripped via mass loss, in the forms of stellar wind for a single-star progenitor or a combination of stellar wind and mass transfer to a companion for a binary progenitor.  Stellar winds depend on metallicity. Simulations of stellar evolution leading to Type IIb SNe have been carried out by \citet{Sravan2019} for metallicities of 1 and 1/4 times the solar value, and by \citet{Claeys2011} for a solar metallicity.  From the former we learn that the single-star path to Type IIb SNe requires higher progenitor masses and that the binary path to Type IIb SNe is dominant at low metallicities.  From the latter we learn that in 90\% cases the secondary star is a hot O star and in 3\% cases the secondary star is an overluminous B star due to accretion, and that a secondary star of roughly equal mass to the SN progenitor would be a K supergiant.

The SMC has a low metallicity, and thus the progenitor of \E\ is most likely in a binary system and the progenitor mass is most likely in the mass range of 10-25 $M_\odot$.  From the photometric and spectroscopic data of stars within \E, we find the earliest types of stars within the boundary of \E\ are B1.5\,V (star 34b) and B3\,V (star 34a).  This is confirmed by the catalog of massive OB stars in the SMC \citep{Massey2002}, which reports OB stars down to B1\,V but shows no entry of OB stars within 60$''$ of \E.  The O stars that are closest to \E\ are in the \ion{He}{3} region LH$\alpha$115-N76 \citep{Naze2003}, whose northeast rim is at 0\farcm8 southwest of E0102.
It is not possible to obscure a luminous O star companion by dust because the photoionization of the condensible species in the SN ejecta suppresses dust formation \citep{Kochanek2017}. 
Thus, we can definitively exclude an O star companion to the SN progenitor of E0102.  

Assuming that the NS reported by \citet{Vogt2018} is real and the SN explosion site is near the NS, the most massive stars within the SNR are stars 34a and 34b.  Both of these stars are at $\sim$4$''$ from the explosion site and would need $\sim$675 km~s$^{-1}$ transverse velocity to run to their present positions.  This corresponds to a proper motion of 2.3 mas yr$^{-1}$, which should have been detected by Gaia EDR3 at a 10$\sigma$ level but was not detected. Therefore, stars 34a and 34b cannot be the runaway companions of the NS's progenitor.  Star 47, with a spectral type of B5\,V and mass of $\sim$ 6 $M_\odot$, is the next most massive star.  Its distance to the NS requires a proper motion of 570 km s$^{-1}$, corresponding to 2.9 mas yr$^{-1}$.  This should be detected at a 4--5$\sigma$ level in Gaia EDR3, but was not detected (see Table \ref{table:gaia}).  The radial velocity of star 47 is only $\sim$50 km s$^{-1}$ relative to the SMC.  There is no evidence supporting star 47 to be a runaway star. The other stars are of later types and smaller masses, and are not in the expected range of binary companions to a 10-25 $M_\odot$ primary star. We conclude that no binary companion of the SN progenitor of \E\ is detected in our search, if the explosion site is near the NS as suggested by \citet{Vogt2018}.

The above considerations assume that the NS was produced in the SN explosion and is near the explosion site.
Since the X-ray point source of the NS reported by \citet{Vogt2018} has been suggested to be an ejecta knot \citep{Long2020}, and since both \citet{Finkelstein2006} and \citet{Banovetz2021} derived an expansion center of the oxygen-rich SN ejecta at $\sim$6$''$ from Vogt et al.'s explosion site, we have repeated the search for surviving companion of the SN progenitor for the explosion site suggested by \citet{Banovetz2021}.  The B3\,V star 34a is the nearest to this explosion center, and has an intriguing spectral bump at 5000--5300 \AA. Its required transverse velocity is 445 $\pm$ 45 km s$^{-1}$, which is high but cannot be immediately excluded because of the uncertainties in the explosion center and the SNR's age.  We conclude that star 34a may be a plausible candidate for the binary companion of E0102's SN progenitor for Banovetz et al.'s site of SN explosion.

We have considered the possibility of a less luminous companion of E0102's SN progenitor, where a significant amount of dust form and obscures the companion \citep{Kochanek2017}.  If a companion is enshrouded in dust, its luminosity will go into longer wavelengths and present large IR excess. We have examined the IRSF catalog \citep{Kato2007} of stars in E0102, and do not see any star with obvious excess in the K band. We do not see any bright source in the Spitzer MIPS 24 $\mu$m band, either. So, we do not have any evidence of a candidate companion enshrouded by dust.

\subsection{What about the Neutron Star?}

The explosion center suggested by \citet{Vogt2018} is based on the presence of a torus of oxygen-rich and neon-rich
material centered near a NS identified in X-ray.  The scenario of star 34a being a companion of the E0102's SN progenitor
for Banovetz et al.'s explosion center does not explain the presence of the torus of metal-rich material or the NS.
The NS can be the product of the SN explosion that produce the SNR
E0102 or it can be a binary companion of the SN's massive progenitor.

The distance of E0102's NS to the explosion centers suggested by \citet{Finkelstein2006} and 
\citet{Banovetz2021} is $\sim6''$.
The NS would need a kick velocity of $\sim$1000 km s$^{-1}$ to reach its current position.  
Such a high kick velocity can be produced in an asymmetric explosion, and some NSs in SNRs have been 
observed to have such high velocities, e.g., N49 in the LMC \citep{Katsuda2018} and the oxygen-rich SNR Puppis A
in the Galaxy \citep{Mayer2020}.  

On the other hand, the NS may be a binary companion of the massive star that has exploded to produce the SNR E0102. 
In this case, the NS was kicked by the SN explosion of its massive companion.
Assuming that the NS suggested by \citet{Vogt2018} was the SN progenitor's companion, it would need a runaway velocity of $\sim1000$ km s$^{-1}$ to reach its present location $\sim$6$''$ from the explosion center of \citet{Banovetz2021}. 
As mentioned above, such high kick velocities have been observed for the NSs in N49 and Puppis A. 
To probe the condition to produce such a high runaway velocity, we assume the SN progenitor was a 15 $M_\odot$ star and the companion is a 1.5 $M_\odot$ NS.  For the NS to have an orbital velocity of 1000 km s$^{-1}$, the orbital radius is about 3.2 $R_\odot$, well within the radius of a 15 $M_\odot$ SN progenitor. For a NS to be inside a SN progenitor, the binary system probably have initially gone through a common envelope phase and the secondary had spiraled toward the primary and remain bound to the NS produced by the primary.
The Secondary had gained mass from the primary and eventually exploded to produce the SNR \E, while the NS had accreted stellar material from the SN progenitor to form the torus of oxygen- and neon-rich material reported by \citet{Vogt2018}. 
This is clearly an over-simplified scenario.
Numerical simulations of such close binary evolution may verify or reject such process.  We encourage modelers to explore the close binary parameter space more finely, especially between the cases where a compact remnant becomes unbound after the first explosion and the cases where the compact remnant from the first explosion and the binary companion merge (e.g.,  \citealt{Renzo2019, Sravan2020}).

Finally, if the NS suggested by \citet{Vogt2018} is indeed an ejecta knot as suggested by \citet{Long2020}, then the 
small torus of oxygen- and neon-rich material detected by \citet{Vogt2018} cannot be explained easily.  It is possible that \E's SN progenitor was a single star or has a low-mass stellar companion; however, the torus of oxygen-rich and neon-rich material at a location with such a large offset from the SN explosion site is also difficult to explain.

\section{Summary}

The young SMC SNR \E\ contains oxygen-rich SN ejecta and a possible NS.  
It also shows a small amount of fast-moving H-rich ejecta material.  
These properties are also observed in Cas A, \E\ has thus been suggested to 
be a Type IIb SNR as in the case of Cas A.  To probe the nature of \E, 
we carried out a search for a surviving binary companion of its SN progenitor.

We have used HST continuum images to make photometric measurements of 
stars in and near \E\ and constructed CMDs to examine the stellar population.
We have also used the VLT MUSE observations to extract spectra of
stars and search for peculiar radial velocities as diagnostics of 
surviving companions. In addition, we have used HST and Gaia data to 
examine proper motions of stars for complementary kinetic studies.  
Using the above data and the massive star catalog of \citet{Massey2002},
we find no MS stars earlier than B1.5 within 60$''$ (18 pc) from \E.  
These results imply that the SN progenitor of E0102 cannot have had an
O star companion.

We first consider the explosion center near the NS, as suggested by 
\citet{Vogt2018}, and examine stars within 9\farcs0 of the explosion
center to search for binary companion candidates of the SN progenitor.
Only two stars have masses in the likely range for 
binary companions of Type IIb SN progenitors: star 34a (B3\,V) and 
34b (B1.5\,V).  These two stars need to have runaway velocities greater 
than 675 km s$^{-1}$ to reach their current positions, but Gaia data 
do not show such 
large proper motions.  Therefore, we do not consider 34a and 34b likely
candidates as a companion of the SN progenitor.  The other stars have
later spectral types with stellar masses too small compared with 
expectations from model predictions.  None of the stars show peculiar 
radial velocities greater than 50 km s$^{-1}$.  
Therefore, no plausible B star candidates are found for a companion of 
E0102's SN progenitor, if the explosion center is near the NS. 

We then consider the explosion center determined from the expansion of
the SN ejecta in \E\ by \citet{Banovetz2021}.  Star 34a is the nearest
MS B star to the explosion center.  Its B3\,V spectral type implies a
mass of 7.6 $M_\odot$, which is within the range of expected companion
masses. Its 2\farcs63 distance to the explosion center requires a transverse
velocity of $445 \pm 45$ km s$^{-1}$, which is high but should not be dismissed 
due to the uncertainty in the explosion center and SN age.  Furthermore, 
The spectrum of star 34a shows a spectral bump in the 5000--5300 \AA\ range,
which may be very interesting if confirmed.
Thus, star 34a may be a plausible candidate as the surviving companion of the
SN progenitor, if the explosion center is as suggested by Banovetz et al.

Finally, if the NS is real and is associated with E0102, it may be the 
product of the SN or a companion of the SN progenitor.  Adopting the 
explosion center determined from the expansion of the SN ejecta by 
\citet{Banovetz2021}, the NS would have a runaway velocity of 
$\sim$1000 km s$^{-1}$.  NSs with such high velocities have been observed  
in the LMC SNR N49 \citep{Katsuda2018} and the Galactic SNR Puppis A \citep{Mayer2020}, and can be achieved
in an asymmetric explosion of a SN that produced the NS, or kicked 
by the explosion of its massive companion.
In either case, the torus of oxygen- and neon-rich material needs to be 
explained, and numerical simulations may help us confirm or reject the
above scenarios.

\acknowledgments
We thank the referee for making helpful suggestions to improve this paper.
Y.-H.C. and C.-J.L. are supported by Taiwanese Ministry of Science and Technology grants MOST 108-2112-M-001-045, 108-2811-M-001-587, 109-2112-M-001-040, and 109-2811-M-001-545. I.R.S. acknowledges support from the Australian Research Council Grant FT160100028. A.J.R. acknowledges support from the Australian Research Council Grant FT170100243.

\emph{Software:} SAOImage DS9 \citep{Joye2003}, QFitsView \citep{Ott2012}, astropy \citep{Astropy2018}, matplotlib \citep{Hunter2007}, numpy \citep{vanderWalt2011, Harris2020}, scipy \citep{Virtanen2020}

\appendix

\section{Appendix A}

\label{appendix:photometry}

We have used HST F475W, F550M, and F775W images of \E\
to study the underlying stellar population. We use 
HST photometrc measurements of stars projected in the 
SNR and its vicinity to construct color-magnitude diagrams 
(CMDs) and compare positions of stars in the CMDs with 
expected surviving companions of Type IIb SNe. The photometrc measurements of stars in our consideration within $\sim$ 9\farcs0 from 
Vogt et al. and Banovetz et al. explosion centers are listed in Table 
{\ref{table:photometryGreen}} and \ref{table:photometryNew}, respectively. In Table \ref{table:photometryNew}, we do not provide new running numbers in order to avoid confusion; instead, running numbers from 
Table~\ref{table:photometryGreen} are used for the few stars that are discussed.


\LongTables
\begin{deluxetable*}{ccccccccc}
\tablecolumns{9}
\tabletypesize{\scriptsize}
\tablewidth{0pc}
\tablecaption{ Stars with F550M$<26$ and within 9\farcs0 from Vogt et al. explosion center}
\tablehead{Star & R.A. (J2000) & Decl. (J2000) & F475W & F550M & F775W & F475W-F550M & F550M-F775W & $r$\\
& & & & & & & & \\}
\startdata
1 & 01:04:02.74 & -72:02:00.93 & 24.19 $\pm$ 0.02 & 23.73 $\pm$ 0.03 & 23.16 $\pm$ 0.02 & 0.46 $\pm$ 0.04 & 0.57 $\pm$ 0.04 & 0.75\arcsec \\ 
2 & 01:04:02.88 & -72:01:59.96 & 27.24 $\pm$ 0.14 & 26.02 $\pm$ 0.12 & 25.34 $\pm$ 0.07 & 1.23 $\pm$ 0.18 & 0.67 $\pm$ 0.14 & 0.88\arcsec \\ 
3 & 01:04:02.73 & -72:02:01.16 & 24.84 $\pm$ 0.02 & 24.24 $\pm$ 0.03 & 23.41 $\pm$ 0.02 & 0.60 $\pm$ 0.04 & 0.82 $\pm$ 0.04 & 0.97\arcsec \\ 
4 & 01:04:02.61 & -72:02:01.46 & 24.49 $\pm$ 0.02 & 24.05 $\pm$ 0.03 & 23.53 $\pm$ 0.02 & 0.44 $\pm$ 0.04 & 0.52 $\pm$ 0.04 & 1.32\arcsec \\ 
5 & 01:04:02.99 & -72:02:00.76 & 25.32 $\pm$ 0.03 & 25.09 $\pm$ 0.06 & 24.25 $\pm$ 0.03 & 0.23 $\pm$ 0.07 & 0.84 $\pm$ 0.07 & 1.46\arcsec \\ 
6 & 01:04:03.04 & -72:02:00.30 & 19.81 $\pm$ 0.00 & 19.78 $\pm$ 0.00 & 19.79 $\pm$ 0.00 & 0.03 $\pm$ 0.00 & -0.01 $\pm$ 0.00 & 1.58\arcsec \\ 
7 & 01:04:02.32 & -72:01:59.94 & 24.36 $\pm$ 0.02 & 23.87 $\pm$ 0.03 & 23.16 $\pm$ 0.02 & 0.49 $\pm$ 0.04 & 0.70 $\pm$ 0.04 & 1.77\arcsec \\ 
8 & 01:04:02.38 & -72:02:01.37 & 23.09 $\pm$ 0.01 & 22.66 $\pm$ 0.01 & 22.08 $\pm$ 0.01 & 0.43 $\pm$ 0.01 & 0.57 $\pm$ 0.01 & 1.88\arcsec \\ 
9 & 01:04:03.11 & -72:02:00.28 & 25.13 $\pm$ 0.03 & 24.81 $\pm$ 0.05 & 24.07 $\pm$ 0.03 & 0.31 $\pm$ 0.06 & 0.74 $\pm$ 0.06 & 1.88\arcsec \\ 
10 & 01:04:03.14 & -72:02:00.51 & 25.95 $\pm$ 0.06 & 25.24 $\pm$ 0.07 & 24.63 $\pm$ 0.04 & 0.71 $\pm$ 0.09 & 0.61 $\pm$ 0.08 & 2.06\arcsec \\ 
11 & 01:04:03.06 & -72:01:58.94 & 24.92 $\pm$ 0.04 & 24.46 $\pm$ 0.05 & 23.87 $\pm$ 0.03 & 0.46 $\pm$ 0.06 & 0.59 $\pm$ 0.06 & 2.10\arcsec \\ 
12 & 01:04:03.13 & -72:02:00.88 & 25.01 $\pm$ 0.03 & 24.53 $\pm$ 0.04 & 23.86 $\pm$ 0.02 & 0.48 $\pm$ 0.05 & 0.68 $\pm$ 0.04 & 2.10\arcsec \\ 
13 & 01:04:02.79 & -72:01:58.09 & 24.03 $\pm$ 0.02 & 23.54 $\pm$ 0.03 & 23.09 $\pm$ 0.02 & 0.49 $\pm$ 0.04 & 0.45 $\pm$ 0.04 & 2.15\arcsec \\ 
14 & 01:04:02.42 & -72:02:02.09 & 26.50 $\pm$ 0.08 & 25.51 $\pm$ 0.08 & 24.83 $\pm$ 0.05 & 0.99 $\pm$ 0.11 & 0.68 $\pm$ 0.09 & 2.29\arcsec \\ 
15 & 01:04:03.21 & -72:02:00.33 & 26.24 $\pm$ 0.07 & 25.52 $\pm$ 0.08 & 24.79 $\pm$ 0.05 & 0.71 $\pm$ 0.11 & 0.73 $\pm$ 0.09 & 2.37\arcsec \\ 
16 & 01:04:02.85 & -72:02:02.65 & 25.95 $\pm$ 0.05 & 25.31 $\pm$ 0.07 & 24.60 $\pm$ 0.04 & 0.64 $\pm$ 0.09 & 0.71 $\pm$ 0.08 & 2.55\arcsec \\ 
17 & 01:04:03.22 & -72:02:01.21 & 23.68 $\pm$ 0.01 & 23.36 $\pm$ 0.02 & 22.79 $\pm$ 0.01 & 0.32 $\pm$ 0.02 & 0.57 $\pm$ 0.02 & 2.61\arcsec \\ 
18 & 01:04:02.47 & -72:02:02.73 & 23.09 $\pm$ 0.01 & 22.67 $\pm$ 0.01 & 22.25 $\pm$ 0.01 & 0.43 $\pm$ 0.01 & 0.42 $\pm$ 0.01 & 2.75\arcsec \\ 
19 & 01:04:02.17 & -72:01:58.91 & 26.01 $\pm$ 0.06 & 25.28 $\pm$ 0.07 & 24.49 $\pm$ 0.04 & 0.74 $\pm$ 0.09 & 0.79 $\pm$ 0.08 & 2.75\arcsec \\ 
20 & 01:04:03.29 & -72:02:00.84 & 26.06 $\pm$ 0.06 & 26.09 $\pm$ 0.13 & 25.07 $\pm$ 0.06 & -0.03 $\pm$ 0.14 & 1.01 $\pm$ 0.14 & 2.83\arcsec \\ 
21 & 01:04:03.19 & -72:01:58.46 & 21.26 $\pm$ 0.01 & 21.21 $\pm$ 0.01 & 21.19 $\pm$ 0.01 & 0.05 $\pm$ 0.01 & 0.02 $\pm$ 0.01 & 2.85\arcsec \\ 
22 & 01:04:03.33 & -72:02:00.15 & 22.40 $\pm$ 0.01 & 22.05 $\pm$ 0.01 & 21.57 $\pm$ 0.01 & 0.35 $\pm$ 0.01 & 0.49 $\pm$ 0.01 & 2.91\arcsec \\ 
23 & 01:04:02.58 & -72:02:03.36 & 25.58 $\pm$ 0.04 & 24.98 $\pm$ 0.06 & 24.34 $\pm$ 0.04 & 0.60 $\pm$ 0.07 & 0.64 $\pm$ 0.07 & 3.20\arcsec \\ 
24 & 01:04:03.40 & -72:02:00.56 & 26.40 $\pm$ 0.08 & 25.97 $\pm$ 0.12 & 24.71 $\pm$ 0.04 & 0.44 $\pm$ 0.14 & 1.25 $\pm$ 0.13 & 3.28\arcsec \\ 
25 & 01:04:02.76 & -72:01:56.93 & 25.16 $\pm$ 0.04 & 24.52 $\pm$ 0.06 & 24.08 $\pm$ 0.04 & 0.65 $\pm$ 0.07 & 0.44 $\pm$ 0.07 & 3.28\arcsec \\ 
26 & 01:04:02.52 & -72:02:03.38 & 22.11 $\pm$ 0.01 & 21.59 $\pm$ 0.01 & 20.92 $\pm$ 0.00 & 0.53 $\pm$ 0.01 & 0.67 $\pm$ 0.01 & 3.29\arcsec \\ 
27 & 01:04:01.98 & -72:02:00.45 & 26.21 $\pm$ 0.07 & 25.49 $\pm$ 0.10 & 24.71 $\pm$ 0.04 & 0.72 $\pm$ 0.12 & 0.78 $\pm$ 0.11 & 3.35\arcsec \\ 
28 & 01:04:02.56 & -72:01:56.81 & 18.86 $\pm$ 0.00 & 18.86 $\pm$ 0.00 & 18.91 $\pm$ 0.00 & 0.00 $\pm$ 0.00 & -0.05 $\pm$ 0.00 & 3.46\arcsec \\ 
29 & 01:04:03.47 & -72:02:00.10 & 23.22 $\pm$ 0.01 & 22.89 $\pm$ 0.01 & 22.42 $\pm$ 0.01 & 0.33 $\pm$ 0.01 & 0.48 $\pm$ 0.01 & 3.57\arcsec \\ 
30 & 01:04:02.14 & -72:02:02.67 & 23.19 $\pm$ 0.01 & 22.80 $\pm$ 0.01 & 22.40 $\pm$ 0.01 & 0.38 $\pm$ 0.01 & 0.41 $\pm$ 0.01 & 3.57\arcsec \\ 
31 & 01:04:02.36 & -72:02:03.55 & 25.53 $\pm$ 0.04 & 25.11 $\pm$ 0.06 & 24.30 $\pm$ 0.03 & 0.43 $\pm$ 0.07 & 0.81 $\pm$ 0.07 & 3.70\arcsec \\ 
32 & 01:04:01.98 & -72:02:01.85 & 25.62 $\pm$ 0.04 & 25.92 $\pm$ 0.12 & 25.93 $\pm$ 0.11 & -0.30 $\pm$ 0.13 & 0.00 $\pm$ 0.16 & 3.73\arcsec \\ 
33 & 01:04:01.90 & -72:01:59.41 & 24.46 $\pm$ 0.02 & 23.99 $\pm$ 0.03 & 23.44 $\pm$ 0.02 & 0.46 $\pm$ 0.04 & 0.55 $\pm$ 0.04 & 3.78\arcsec \\ 
34 & 01:04:03.29 & -72:01:57.56 & 26.61 $\pm$ 0.12 & 25.53 $\pm$ 0.12 & 24.84 $\pm$ 0.07 & 1.08 $\pm$ 0.17 & 0.69 $\pm$ 0.14 & 3.79\arcsec \\ 
35 & 01:04:03.03 & -72:02:03.96 & 25.79 $\pm$ 0.05 & 25.29 $\pm$ 0.07 & 24.43 $\pm$ 0.04 & 0.50 $\pm$ 0.09 & 0.86 $\pm$ 0.08 & 4.06\arcsec \\ 
36 & 01:04:02.25 & -72:01:56.64 & 24.34 $\pm$ 0.03 & 23.93 $\pm$ 0.04 & 23.35 $\pm$ 0.03 & 0.41 $\pm$ 0.05 & 0.58 $\pm$ 0.05 & 4.13\arcsec \\ 
37 & 01:04:02.54 & -72:01:55.91 & 25.96 $\pm$ 0.08 & 25.46 $\pm$ 0.12 & 24.35 $\pm$ 0.06 & 0.50 $\pm$ 0.14 & 1.11 $\pm$ 0.13 & 4.35\arcsec \\ 
38 & 01:04:02.69 & -72:02:04.73 & 22.27 $\pm$ 0.01 & 22.06 $\pm$ 0.01 & 21.80 $\pm$ 0.01 & 0.20 $\pm$ 0.01 & 0.26 $\pm$ 0.01 & 4.53\arcsec \\ 
39 & 01:04:03.04 & -72:01:55.95 & 23.29 $\pm$ 0.01 & 22.95 $\pm$ 0.02 & 22.48 $\pm$ 0.02 & 0.34 $\pm$ 0.02 & 0.47 $\pm$ 0.03 & 4.53\arcsec \\ 
40 & 01:04:02.08 & -72:01:56.69 & 22.06 $\pm$ 0.01 & 21.87 $\pm$ 0.01 & 21.67 $\pm$ 0.01 & 0.18 $\pm$ 0.01 & 0.20 $\pm$ 0.01 & 4.54\arcsec \\ 
41 & 01:04:01.98 & -72:02:03.56 & 26.55 $\pm$ 0.09 & 25.83 $\pm$ 0.11 & 25.13 $\pm$ 0.06 & 0.71 $\pm$ 0.14 & 0.70 $\pm$ 0.13 & 4.73\arcsec \\ 
42 & 01:04:01.94 & -72:02:03.44 & 22.83 $\pm$ 0.01 & 22.55 $\pm$ 0.01 & 22.21 $\pm$ 0.01 & 0.28 $\pm$ 0.01 & 0.33 $\pm$ 0.01 & 4.80\arcsec \\ 
43 & 01:04:02.66 & -72:02:05.01 & 25.30 $\pm$ 0.03 & 24.74 $\pm$ 0.05 & 24.19 $\pm$ 0.03 & 0.56 $\pm$ 0.06 & 0.55 $\pm$ 0.06 & 4.81\arcsec \\ 
44 & 01:04:03.37 & -72:01:56.29 & 25.33 $\pm$ 0.03 & 24.53 $\pm$ 0.04 & 23.62 $\pm$ 0.02 & 0.80 $\pm$ 0.05 & 0.91 $\pm$ 0.04 & 4.98\arcsec \\ 
45 & 01:04:02.90 & -72:01:55.28 & 24.44 $\pm$ 0.03 & 24.13 $\pm$ 0.04 & 23.50 $\pm$ 0.03 & 0.31 $\pm$ 0.05 & 0.63 $\pm$ 0.05 & 5.01\arcsec \\ 
46 & 01:04:03.13 & -72:01:55.56 & 26.17 $\pm$ 0.10 & 25.78 $\pm$ 0.14 & 24.93 $\pm$ 0.07 & 0.38 $\pm$ 0.17 & 0.85 $\pm$ 0.16 & 5.05\arcsec \\ 
47 & 01:04:02.11 & -72:02:04.48 & 20.04 $\pm$ 0.00 & 20.05 $\pm$ 0.00 & 20.08 $\pm$ 0.00 & -0.01 $\pm$ 0.00 & -0.03 $\pm$ 0.00 & 5.08\arcsec \\ 
48 & 01:04:03.66 & -72:01:57.69 & 26.12 $\pm$ 0.08 & 25.56 $\pm$ 0.12 & 24.57 $\pm$ 0.06 & 0.56 $\pm$ 0.14 & 0.99 $\pm$ 0.13 & 5.09\arcsec \\ 
49 & 01:04:03.75 & -72:02:01.80 & 26.08 $\pm$ 0.06 & 25.29 $\pm$ 0.07 & 24.69 $\pm$ 0.04 & 0.80 $\pm$ 0.09 & 0.60 $\pm$ 0.08 & 5.12\arcsec \\ 
50 & 01:04:03.48 & -72:01:56.50 & 25.68 $\pm$ 0.06 & 25.06 $\pm$ 0.08 & 24.31 $\pm$ 0.05 & 0.62 $\pm$ 0.10 & 0.75 $\pm$ 0.09 & 5.17\arcsec \\ 
51 & 01:04:01.57 & -72:01:59.43 & 21.67 $\pm$ 0.00 & 21.52 $\pm$ 0.01 & 21.38 $\pm$ 0.01 & 0.16 $\pm$ 0.01 & 0.13 $\pm$ 0.01 & 5.30\arcsec \\ 
52 & 01:04:03.57 & -72:02:03.67 & 22.48 $\pm$ 0.01 & 22.20 $\pm$ 0.01 & 21.76 $\pm$ 0.01 & 0.27 $\pm$ 0.01 & 0.45 $\pm$ 0.01 & 5.30\arcsec \\ 
53 & 01:04:01.89 & -72:02:04.05 & 23.14 $\pm$ 0.01 & 22.77 $\pm$ 0.02 & 22.32 $\pm$ 0.01 & 0.37 $\pm$ 0.02 & 0.45 $\pm$ 0.02 & 5.36\arcsec \\ 
54 & 01:04:01.53 & -72:02:00.50 & 25.93 $\pm$ 0.06 & 25.70 $\pm$ 0.10 & 24.74 $\pm$ 0.05 & 0.23 $\pm$ 0.12 & 0.96 $\pm$ 0.11 & 5.42\arcsec \\ 
55 & 01:04:01.76 & -72:01:56.92 & 22.66 $\pm$ 0.01 & 22.35 $\pm$ 0.02 & 21.95 $\pm$ 0.01 & 0.31 $\pm$ 0.02 & 0.40 $\pm$ 0.02 & 5.45\arcsec \\ 
56 & 01:04:03.48 & -72:02:04.32 & 26.22 $\pm$ 0.06 & 25.54 $\pm$ 0.08 & 24.74 $\pm$ 0.05 & 0.68 $\pm$ 0.10 & 0.80 $\pm$ 0.09 & 5.47\arcsec \\ 
57 & 01:04:03.75 & -72:02:02.92 & 25.31 $\pm$ 0.03 & 24.72 $\pm$ 0.05 & 24.00 $\pm$ 0.03 & 0.59 $\pm$ 0.06 & 0.72 $\pm$ 0.06 & 5.57\arcsec \\ 
58 & 01:04:03.79 & -72:02:02.55 & 24.38 $\pm$ 0.02 & 23.60 $\pm$ 0.02 & 22.54 $\pm$ 0.01 & 0.78 $\pm$ 0.03 & 1.06 $\pm$ 0.02 & 5.57\arcsec \\ 
59 & 01:04:03.08 & -72:02:05.59 & 24.02 $\pm$ 0.01 & 23.67 $\pm$ 0.02 & 23.12 $\pm$ 0.02 & 0.35 $\pm$ 0.02 & 0.55 $\pm$ 0.03 & 5.68\arcsec \\ 
60 & 01:04:01.95 & -72:02:04.86 & 26.25 $\pm$ 0.07 & 25.22 $\pm$ 0.07 & 24.54 $\pm$ 0.05 & 1.02 $\pm$ 0.10 & 0.68 $\pm$ 0.09 & 5.82\arcsec \\ 
61 & 01:04:02.40 & -72:01:54.50 & 25.21 $\pm$ 0.04 & 24.56 $\pm$ 0.06 & 24.00 $\pm$ 0.04 & 0.65 $\pm$ 0.07 & 0.56 $\pm$ 0.07 & 5.86\arcsec \\ 
62 & 01:04:03.25 & -72:01:54.82 & 21.65 $\pm$ 0.01 & 21.47 $\pm$ 0.01 & 21.22 $\pm$ 0.01 & 0.18 $\pm$ 0.01 & 0.25 $\pm$ 0.01 & 5.96\arcsec \\ 
63 & 01:04:01.52 & -72:01:57.63 & 26.48 $\pm$ 0.16 & 25.90 $\pm$ 0.21 & 25.18 $\pm$ 0.13 & 0.58 $\pm$ 0.26 & 0.72 $\pm$ 0.25 & 6.03\arcsec \\ 
64 & 01:04:01.42 & -72:01:59.06 & 25.51 $\pm$ 0.04 & 25.15 $\pm$ 0.06 & 24.40 $\pm$ 0.04 & 0.36 $\pm$ 0.07 & 0.75 $\pm$ 0.07 & 6.03\arcsec \\ 
65 & 01:04:02.55 & -72:02:06.21 & 25.03 $\pm$ 0.03 & 24.44 $\pm$ 0.04 & 23.74 $\pm$ 0.02 & 0.58 $\pm$ 0.05 & 0.71 $\pm$ 0.04 & 6.05\arcsec \\ 
66 & 01:04:03.98 & -72:02:01.57 & 26.58 $\pm$ 0.10 & 25.88 $\pm$ 0.11 & 24.93 $\pm$ 0.05 & 0.70 $\pm$ 0.15 & 0.95 $\pm$ 0.12 & 6.07\arcsec \\ 
67 & 01:04:03.84 & -72:02:03.31 & 26.79 $\pm$ 0.11 & 25.69 $\pm$ 0.10 & 24.75 $\pm$ 0.05 & 1.10 $\pm$ 0.15 & 0.95 $\pm$ 0.11 & 6.13\arcsec \\ 
68 & 01:04:02.76 & -72:01:54.02 & 24.26 $\pm$ 0.02 & 23.82 $\pm$ 0.03 & 23.17 $\pm$ 0.02 & 0.45 $\pm$ 0.04 & 0.64 $\pm$ 0.04 & 6.18\arcsec \\ 
69 & 01:04:03.43 & -72:01:55.02 & 24.64 $\pm$ 0.03 & 24.17 $\pm$ 0.05 & 23.61 $\pm$ 0.03 & 0.46 $\pm$ 0.06 & 0.57 $\pm$ 0.06 & 6.19\arcsec \\ 
70 & 01:04:01.36 & -72:02:00.17 & 24.77 $\pm$ 0.03 & 24.02 $\pm$ 0.04 & 23.12 $\pm$ 0.02 & 0.75 $\pm$ 0.05 & 0.90 $\pm$ 0.04 & 6.21\arcsec \\ 
71 & 01:04:03.87 & -72:02:03.34 & 25.57 $\pm$ 0.04 & 24.96 $\pm$ 0.05 & 24.32 $\pm$ 0.03 & 0.61 $\pm$ 0.06 & 0.64 $\pm$ 0.06 & 6.27\arcsec \\ 
72 & 01:04:03.20 & -72:02:06.10 & 26.59 $\pm$ 0.08 & 25.78 $\pm$ 0.10 & 24.85 $\pm$ 0.05 & 0.81 $\pm$ 0.13 & 0.93 $\pm$ 0.11 & 6.33\arcsec \\ 
73 & 01:04:03.16 & -72:01:54.19 & 24.64 $\pm$ 0.03 & 24.19 $\pm$ 0.04 & 23.63 $\pm$ 0.03 & 0.45 $\pm$ 0.05 & 0.56 $\pm$ 0.05 & 6.38\arcsec \\ 
74 & 01:04:04.01 & -72:01:58.18 & 25.09 $\pm$ 0.04 & 24.44 $\pm$ 0.05 & 23.78 $\pm$ 0.03 & 0.65 $\pm$ 0.06 & 0.66 $\pm$ 0.06 & 6.38\arcsec \\ 
75 & 01:04:03.08 & -72:02:06.38 & 25.85 $\pm$ 0.05 & 25.11 $\pm$ 0.06 & 24.54 $\pm$ 0.04 & 0.74 $\pm$ 0.08 & 0.57 $\pm$ 0.07 & 6.42\arcsec \\ 
76 & 01:04:01.68 & -72:01:55.81 & 24.74 $\pm$ 0.03 & 24.60 $\pm$ 0.06 & 23.95 $\pm$ 0.04 & 0.14 $\pm$ 0.07 & 0.65 $\pm$ 0.07 & 6.45\arcsec \\ 
77 & 01:04:03.87 & -72:02:03.82 & 25.08 $\pm$ 0.03 & 24.62 $\pm$ 0.04 & 23.89 $\pm$ 0.03 & 0.46 $\pm$ 0.05 & 0.72 $\pm$ 0.05 & 6.51\arcsec \\ 
78 & 01:04:01.41 & -72:01:57.59 & 26.54 $\pm$ 0.13 & 26.02 $\pm$ 0.19 & 25.48 $\pm$ 0.12 & 0.53 $\pm$ 0.23 & 0.54 $\pm$ 0.22 & 6.52\arcsec \\ 
79 & 01:04:01.52 & -72:02:03.88 & 26.44 $\pm$ 0.08 & 25.89 $\pm$ 0.11 & 24.99 $\pm$ 0.05 & 0.55 $\pm$ 0.14 & 0.90 $\pm$ 0.12 & 6.58\arcsec \\ 
80 & 01:04:02.92 & -72:02:06.71 & 21.48 $\pm$ 0.00 & 21.04 $\pm$ 0.01 & 20.41 $\pm$ 0.00 & 0.44 $\pm$ 0.01 & 0.63 $\pm$ 0.01 & 6.59\arcsec \\ 
81 & 01:04:03.82 & -72:01:56.09 & 25.07 $\pm$ 0.04 & 24.69 $\pm$ 0.06 & 23.90 $\pm$ 0.04 & 0.38 $\pm$ 0.07 & 0.80 $\pm$ 0.07 & 6.60\arcsec \\ 
82 & 01:04:04.03 & -72:01:57.76 & 21.83 $\pm$ 0.01 & 21.68 $\pm$ 0.01 & 21.45 $\pm$ 0.01 & 0.16 $\pm$ 0.01 & 0.23 $\pm$ 0.01 & 6.62\arcsec \\ 
83 & 01:04:03.91 & -72:02:03.74 & 24.11 $\pm$ 0.02 & 23.64 $\pm$ 0.02 & 23.06 $\pm$ 0.02 & 0.47 $\pm$ 0.03 & 0.59 $\pm$ 0.03 & 6.64\arcsec \\ 
84 & 01:04:01.28 & -72:02:01.13 & 24.33 $\pm$ 0.02 & 23.82 $\pm$ 0.03 & 23.33 $\pm$ 0.02 & 0.51 $\pm$ 0.04 & 0.49 $\pm$ 0.04 & 6.65\arcsec \\ 
85 & 01:04:01.33 & -72:02:02.44 & 26.18 $\pm$ 0.07 & 25.86 $\pm$ 0.11 & 24.88 $\pm$ 0.06 & 0.33 $\pm$ 0.13 & 0.98 $\pm$ 0.13 & 6.73\arcsec \\ 
86 & 01:04:02.38 & -72:02:06.77 & 24.12 $\pm$ 0.02 & 23.73 $\pm$ 0.02 & 23.22 $\pm$ 0.02 & 0.40 $\pm$ 0.03 & 0.51 $\pm$ 0.03 & 6.74\arcsec \\ 
87 & 01:04:01.34 & -72:01:57.58 & 26.13 $\pm$ 0.09 & 25.19 $\pm$ 0.10 & 23.78 $\pm$ 0.04 & 0.94 $\pm$ 0.13 & 1.41 $\pm$ 0.11 & 6.80\arcsec \\ 
88 & 01:04:03.72 & -72:02:05.11 & 23.40 $\pm$ 0.01 & 23.01 $\pm$ 0.02 & 22.51 $\pm$ 0.01 & 0.39 $\pm$ 0.02 & 0.50 $\pm$ 0.02 & 6.81\arcsec \\ 
89 & 01:04:02.04 & -72:02:06.30 & 27.01 $\pm$ 0.12 & 25.95 $\pm$ 0.13 & 24.45 $\pm$ 0.04 & 1.06 $\pm$ 0.18 & 1.50 $\pm$ 0.14 & 6.82\arcsec \\ 
90 & 01:04:01.23 & -72:01:59.59 & 24.14 $\pm$ 0.02 & 23.74 $\pm$ 0.02 & 23.16 $\pm$ 0.02 & 0.41 $\pm$ 0.03 & 0.57 $\pm$ 0.03 & 6.84\arcsec \\ 
91 & 01:04:03.49 & -72:02:05.99 & 26.02 $\pm$ 0.05 & 25.90 $\pm$ 0.11 & 24.76 $\pm$ 0.05 & 0.13 $\pm$ 0.12 & 1.13 $\pm$ 0.12 & 6.84\arcsec \\ 
92 & 01:04:03.81 & -72:01:55.65 & 26.09 $\pm$ 0.09 & 25.31 $\pm$ 0.10 & 24.26 $\pm$ 0.05 & 0.78 $\pm$ 0.13 & 1.06 $\pm$ 0.11 & 6.87\arcsec \\ 
93 & 01:04:01.35 & -72:01:57.28 & 26.55 $\pm$ 0.12 & 25.05 $\pm$ 0.09 & 24.52 $\pm$ 0.06 & 1.50 $\pm$ 0.15 & 0.53 $\pm$ 0.11 & 6.89\arcsec \\ 
94 & 01:04:04.13 & -72:02:02.12 & 25.99 $\pm$ 0.06 & 25.29 $\pm$ 0.07 & 23.90 $\pm$ 0.03 & 0.70 $\pm$ 0.09 & 1.39 $\pm$ 0.08 & 6.90\arcsec \\ 
95 & 01:04:04.20 & -72:02:01.24 & 24.91 $\pm$ 0.03 & 24.36 $\pm$ 0.04 & 23.83 $\pm$ 0.02 & 0.55 $\pm$ 0.05 & 0.53 $\pm$ 0.04 & 7.03\arcsec \\ 
96 & 01:04:01.20 & -72:01:58.85 & 26.40 $\pm$ 0.08 & 26.01 $\pm$ 0.14 & 25.93 $\pm$ 0.12 & 0.39 $\pm$ 0.16 & 0.08 $\pm$ 0.18 & 7.07\arcsec \\ 
97 & 01:04:02.39 & -72:02:07.14 & 26.50 $\pm$ 0.08 & 25.66 $\pm$ 0.09 & 24.28 $\pm$ 0.03 & 0.84 $\pm$ 0.12 & 1.39 $\pm$ 0.09 & 7.08\arcsec \\ 
98 & 01:04:01.20 & -72:01:58.70 & 26.34 $\pm$ 0.08 & 25.48 $\pm$ 0.08 & 25.89 $\pm$ 0.12 & 0.85 $\pm$ 0.11 & -0.41 $\pm$ 0.14 & 7.08\arcsec \\ 
99 & 01:04:01.87 & -72:02:06.23 & 25.85 $\pm$ 0.07 & 25.10 $\pm$ 0.08 & 24.44 $\pm$ 0.05 & 0.75 $\pm$ 0.11 & 0.66 $\pm$ 0.09 & 7.15\arcsec \\ 
100 & 01:04:03.94 & -72:02:04.54 & 24.42 $\pm$ 0.02 & 23.92 $\pm$ 0.03 & 23.32 $\pm$ 0.02 & 0.50 $\pm$ 0.04 & 0.60 $\pm$ 0.04 & 7.20\arcsec \\ 
101 & 01:04:02.60 & -72:01:52.97 & 23.50 $\pm$ 0.02 & 23.28 $\pm$ 0.02 & 22.71 $\pm$ 0.02 & 0.22 $\pm$ 0.03 & 0.57 $\pm$ 0.03 & 7.24\arcsec \\ 
102 & 01:04:03.99 & -72:01:56.05 & 25.94 $\pm$ 0.08 & 25.49 $\pm$ 0.11 & 24.68 $\pm$ 0.06 & 0.45 $\pm$ 0.14 & 0.81 $\pm$ 0.13 & 7.26\arcsec \\ 
103 & 01:04:04.25 & -72:01:58.70 & 25.60 $\pm$ 0.06 & 25.11 $\pm$ 0.09 & 24.13 $\pm$ 0.04 & 0.50 $\pm$ 0.11 & 0.98 $\pm$ 0.10 & 7.35\arcsec \\ 
104 & 01:04:01.80 & -72:01:54.10 & 26.63 $\pm$ 0.13 & 25.82 $\pm$ 0.14 & 24.92 $\pm$ 0.07 & 0.80 $\pm$ 0.19 & 0.90 $\pm$ 0.16 & 7.39\arcsec \\ 
105 & 01:04:02.65 & -72:02:07.59 & 25.74 $\pm$ 0.04 & 25.50 $\pm$ 0.08 & 24.51 $\pm$ 0.04 & 0.24 $\pm$ 0.09 & 0.99 $\pm$ 0.09 & 7.39\arcsec \\ 
106 & 01:04:01.66 & -72:01:54.55 & 20.42 $\pm$ 0.00 & 20.43 $\pm$ 0.01 & 20.45 $\pm$ 0.01 & -0.01 $\pm$ 0.01 & -0.02 $\pm$ 0.01 & 7.42\arcsec \\ 
107 & 01:04:02.67 & -72:01:52.77 & 22.11 $\pm$ 0.01 & 21.83 $\pm$ 0.01 & 21.42 $\pm$ 0.01 & 0.29 $\pm$ 0.01 & 0.41 $\pm$ 0.01 & 7.43\arcsec \\ 
108 & 01:04:02.10 & -72:02:07.16 & 25.09 $\pm$ 0.03 & 24.59 $\pm$ 0.04 & 23.77 $\pm$ 0.02 & 0.50 $\pm$ 0.05 & 0.83 $\pm$ 0.04 & 7.49\arcsec \\ 
109 & 01:04:01.23 & -72:01:56.99 & 25.62 $\pm$ 0.07 & 25.00 $\pm$ 0.09 & 23.66 $\pm$ 0.03 & 0.62 $\pm$ 0.11 & 1.35 $\pm$ 0.09 & 7.51\arcsec \\ 
110 & 01:04:02.12 & -72:01:53.16 & 25.01 $\pm$ 0.04 & 24.53 $\pm$ 0.06 & 23.89 $\pm$ 0.04 & 0.48 $\pm$ 0.07 & 0.64 $\pm$ 0.07 & 7.54\arcsec \\ 
111 & 01:04:01.65 & -72:02:06.00 & 24.02 $\pm$ 0.02 & 23.73 $\pm$ 0.03 & 23.16 $\pm$ 0.02 & 0.29 $\pm$ 0.04 & 0.57 $\pm$ 0.04 & 7.55\arcsec \\ 
112 & 01:04:03.49 & -72:01:53.50 & 27.10 $\pm$ 0.19 & 26.12 $\pm$ 0.19 & 25.43 $\pm$ 0.11 & 0.98 $\pm$ 0.27 & 0.69 $\pm$ 0.22 & 7.62\arcsec \\ 
113 & 01:04:01.63 & -72:02:06.04 & 24.04 $\pm$ 0.02 & 23.62 $\pm$ 0.02 & 23.09 $\pm$ 0.02 & 0.42 $\pm$ 0.03 & 0.54 $\pm$ 0.03 & 7.65\arcsec \\ 
114 & 01:04:01.24 & -72:01:56.56 & 26.60 $\pm$ 0.13 & 26.12 $\pm$ 0.21 & 25.31 $\pm$ 0.11 & 0.48 $\pm$ 0.25 & 0.81 $\pm$ 0.24 & 7.68\arcsec \\ 
115 & 01:04:02.50 & -72:01:52.50 & 25.69 $\pm$ 0.04 & 25.17 $\pm$ 0.06 & 24.50 $\pm$ 0.04 & 0.52 $\pm$ 0.07 & 0.67 $\pm$ 0.07 & 7.75\arcsec \\ 
116 & 01:04:03.70 & -72:01:53.93 & 25.97 $\pm$ 0.07 & 25.36 $\pm$ 0.10 & 24.61 $\pm$ 0.06 & 0.61 $\pm$ 0.12 & 0.76 $\pm$ 0.12 & 7.79\arcsec \\ 
117 & 01:04:03.17 & -72:01:52.69 & 26.96 $\pm$ 0.11 & 26.14 $\pm$ 0.14 & 25.17 $\pm$ 0.06 & 0.83 $\pm$ 0.18 & 0.97 $\pm$ 0.15 & 7.82\arcsec \\ 
118 & 01:04:01.10 & -72:01:57.66 & 25.76 $\pm$ 0.07 & 25.16 $\pm$ 0.10 & 25.20 $\pm$ 0.10 & 0.60 $\pm$ 0.12 & -0.04 $\pm$ 0.14 & 7.82\arcsec \\ 
119 & 01:04:03.79 & -72:02:06.20 & 25.88 $\pm$ 0.05 & 25.42 $\pm$ 0.08 & 24.48 $\pm$ 0.04 & 0.46 $\pm$ 0.09 & 0.94 $\pm$ 0.09 & 7.84\arcsec \\ 
120 & 01:04:03.46 & -72:02:07.24 & 23.95 $\pm$ 0.01 & 23.51 $\pm$ 0.02 & 22.92 $\pm$ 0.01 & 0.43 $\pm$ 0.02 & 0.60 $\pm$ 0.02 & 7.85\arcsec \\ 
121 & 01:04:01.42 & -72:01:55.03 & 22.46 $\pm$ 0.01 & 22.22 $\pm$ 0.01 & 21.85 $\pm$ 0.01 & 0.24 $\pm$ 0.01 & 0.37 $\pm$ 0.01 & 7.88\arcsec \\ 
122 & 01:04:03.49 & -72:01:53.18 & 23.89 $\pm$ 0.01 & 23.50 $\pm$ 0.02 & 22.99 $\pm$ 0.01 & 0.39 $\pm$ 0.02 & 0.51 $\pm$ 0.02 & 7.91\arcsec \\ 
123 & 01:04:02.54 & -72:02:08.08 & 23.68 $\pm$ 0.01 & 23.29 $\pm$ 0.02 & 22.78 $\pm$ 0.01 & 0.39 $\pm$ 0.02 & 0.52 $\pm$ 0.02 & 7.92\arcsec \\ 
124 & 01:04:01.10 & -72:01:57.30 & 25.79 $\pm$ 0.07 & 25.34 $\pm$ 0.11 & 24.82 $\pm$ 0.08 & 0.45 $\pm$ 0.13 & 0.53 $\pm$ 0.14 & 7.95\arcsec \\ 
125 & 01:04:04.37 & -72:01:58.35 & 21.65 $\pm$ 0.01 & 21.46 $\pm$ 0.01 & 21.25 $\pm$ 0.01 & 0.19 $\pm$ 0.01 & 0.21 $\pm$ 0.01 & 7.97\arcsec \\ 
126 & 01:04:01.10 & -72:01:57.16 & 26.29 $\pm$ 0.10 & 25.46 $\pm$ 0.12 & 24.19 $\pm$ 0.05 & 0.84 $\pm$ 0.16 & 1.27 $\pm$ 0.13 & 8.00\arcsec \\ 
127 & 01:04:03.26 & -72:02:07.84 & 25.58 $\pm$ 0.04 & 25.07 $\pm$ 0.06 & 24.40 $\pm$ 0.03 & 0.52 $\pm$ 0.07 & 0.66 $\pm$ 0.07 & 8.07\arcsec \\ 
128 & 01:04:00.97 & -72:01:58.85 & 26.70 $\pm$ 0.10 & 26.03 $\pm$ 0.12 & 24.62 $\pm$ 0.04 & 0.66 $\pm$ 0.16 & 1.42 $\pm$ 0.13 & 8.13\arcsec \\ 
129 & 01:04:04.46 & -72:01:59.65 & 26.84 $\pm$ 0.15 & 26.16 $\pm$ 0.20 & 25.42 $\pm$ 0.11 & 0.68 $\pm$ 0.25 & 0.74 $\pm$ 0.23 & 8.15\arcsec \\ 
130 & 01:04:01.39 & -72:01:54.63 & 26.64 $\pm$ 0.13 & 26.16 $\pm$ 0.19 & 25.03 $\pm$ 0.08 & 0.48 $\pm$ 0.23 & 1.14 $\pm$ 0.21 & 8.24\arcsec \\ 
131 & 01:04:03.64 & -72:02:07.23 & 25.52 $\pm$ 0.04 & 25.00 $\pm$ 0.06 & 24.30 $\pm$ 0.03 & 0.53 $\pm$ 0.07 & 0.69 $\pm$ 0.07 & 8.27\arcsec \\ 
132 & 01:04:03.92 & -72:02:06.48 & 22.27 $\pm$ 0.01 & 22.05 $\pm$ 0.01 & 21.80 $\pm$ 0.01 & 0.23 $\pm$ 0.01 & 0.25 $\pm$ 0.01 & 8.44\arcsec \\ 
133 & 01:04:01.73 & -72:02:07.42 & 24.21 $\pm$ 0.02 & 23.88 $\pm$ 0.03 & 23.32 $\pm$ 0.02 & 0.33 $\pm$ 0.04 & 0.56 $\pm$ 0.04 & 8.51\arcsec \\ 
134 & 01:04:04.17 & -72:02:05.38 & 25.37 $\pm$ 0.04 & 24.87 $\pm$ 0.05 & 24.09 $\pm$ 0.03 & 0.50 $\pm$ 0.06 & 0.78 $\pm$ 0.06 & 8.54\arcsec \\ 
135 & 01:04:01.12 & -72:01:55.80 & 23.82 $\pm$ 0.02 & 23.49 $\pm$ 0.03 & 22.93 $\pm$ 0.02 & 0.33 $\pm$ 0.04 & 0.56 $\pm$ 0.04 & 8.54\arcsec \\ 
136 & 01:04:01.66 & -72:02:07.28 & 25.71 $\pm$ 0.04 & 25.12 $\pm$ 0.06 & 24.45 $\pm$ 0.04 & 0.59 $\pm$ 0.07 & 0.67 $\pm$ 0.07 & 8.56\arcsec \\ 
137 & 01:04:00.86 & -72:01:59.30 & 24.04 $\pm$ 0.01 & 23.60 $\pm$ 0.02 & 23.05 $\pm$ 0.02 & 0.44 $\pm$ 0.02 & 0.55 $\pm$ 0.03 & 8.58\arcsec \\ 
138 & 01:04:03.79 & -72:01:53.25 & 25.88 $\pm$ 0.05 & 25.22 $\pm$ 0.07 & 24.16 $\pm$ 0.03 & 0.66 $\pm$ 0.09 & 1.06 $\pm$ 0.08 & 8.59\arcsec \\ 
139 & 01:04:00.83 & -72:02:00.80 & 26.70 $\pm$ 0.09 & 25.93 $\pm$ 0.12 & 25.16 $\pm$ 0.06 & 0.77 $\pm$ 0.15 & 0.76 $\pm$ 0.13 & 8.66\arcsec \\ 
140 & 01:04:00.84 & -72:02:01.06 & 22.08 $\pm$ 0.01 & 21.67 $\pm$ 0.01 & 21.08 $\pm$ 0.01 & 0.42 $\pm$ 0.01 & 0.58 $\pm$ 0.01 & 8.66\arcsec \\ 
141 & 01:04:01.99 & -72:01:52.14 & 22.18 $\pm$ 0.01 & 22.00 $\pm$ 0.01 & 21.84 $\pm$ 0.01 & 0.19 $\pm$ 0.01 & 0.16 $\pm$ 0.01 & 8.70\arcsec \\ 
142 & 01:04:01.22 & -72:02:05.73 & 25.00 $\pm$ 0.03 & 24.49 $\pm$ 0.04 & 23.87 $\pm$ 0.02 & 0.51 $\pm$ 0.05 & 0.62 $\pm$ 0.04 & 8.82\arcsec \\ 
143 & 01:04:02.54 & -72:01:51.42 & 22.70 $\pm$ 0.01 & 22.46 $\pm$ 0.01 & 22.08 $\pm$ 0.01 & 0.24 $\pm$ 0.01 & 0.38 $\pm$ 0.01 & 8.82\arcsec \\ 
144 & 01:04:00.96 & -72:01:56.38 & 20.53 $\pm$ 0.00 & 20.48 $\pm$ 0.01 & 20.48 $\pm$ 0.01 & 0.04 $\pm$ 0.01 & 0.01 $\pm$ 0.01 & 8.89\arcsec \\ 
145 & 01:04:00.90 & -72:02:03.32 & 26.43 $\pm$ 0.08 & 25.96 $\pm$ 0.12 & 25.08 $\pm$ 0.06 & 0.48 $\pm$ 0.14 & 0.87 $\pm$ 0.13 & 8.90\arcsec \\ 
146 & 01:04:02.66 & -72:02:09.18 & 26.17 $\pm$ 0.06 & 25.56 $\pm$ 0.09 & 24.77 $\pm$ 0.05 & 0.61 $\pm$ 0.11 & 0.78 $\pm$ 0.10 & 8.98\arcsec \\ 
147 & 01:04:03.63 & -72:01:52.32 & 22.18 $\pm$ 0.01 & 21.60 $\pm$ 0.01 & 20.78 $\pm$ 0.00 & 0.57 $\pm$ 0.01 & 0.82 $\pm$ 0.01 & 8.98\arcsec 
\enddata
\label{table:photometryGreen}
\end{deluxetable*}


\LongTables
\begin{deluxetable*}{ccccccccc}
\tablecolumns{9}
\tabletypesize{\scriptsize}
\tablewidth{0pc}
\tablecaption{ Stars with F550M$<24$ and within 9\farcs0 from Banovetz et al. explosion center}
\tablehead{Star & R.A. (J2000) & Decl. (J2000) & F475W & F550M & F775W & F475W-F550M & F550M-F775W & $r$\\
& & & & & & & & \\}
\startdata
- & 01:04:02.60 & -72:01:52.97 & 23.50 $\pm$ 0.02 & 23.28 $\pm$ 0.02 & 22.71 $\pm$ 0.02 & 0.22 $\pm$ 0.03 & 0.57 $\pm$ 0.03 & 1.10\arcsec \\ 
- & 01:04:02.76 & -72:01:54.02 & 24.26 $\pm$ 0.02 & 23.82 $\pm$ 0.03 & 23.17 $\pm$ 0.02 & 0.45 $\pm$ 0.04 & 0.64 $\pm$ 0.04 & 1.29\arcsec \\ 
- & 01:04:02.67 & -72:01:52.77 & 22.11 $\pm$ 0.01 & 21.83 $\pm$ 0.01 & 21.42 $\pm$ 0.01 & 0.29 $\pm$ 0.01 & 0.41 $\pm$ 0.01 & 1.46\arcsec \\ 
- & 01:04:02.54 & -72:01:51.42 & 22.70 $\pm$ 0.01 & 22.46 $\pm$ 0.01 & 22.08 $\pm$ 0.01 & 0.24 $\pm$ 0.01 & 0.38 $\pm$ 0.01 & 2.52\arcsec \\ 
- & 01:04:01.99 & -72:01:52.14 & 22.18 $\pm$ 0.01 & 22.00 $\pm$ 0.01 & 21.84 $\pm$ 0.01 & 0.19 $\pm$ 0.01 & 0.16 $\pm$ 0.01 & 2.87\arcsec \\ 
28 & 01:04:02.56 & -72:01:56.81 & 18.86 $\pm$ 0.00 & 18.86 $\pm$ 0.00 & 18.91 $\pm$ 0.00 & 0.00 $\pm$ 0.00 & -0.05 $\pm$ 0.00 & 2.91\arcsec \\ 
- & 01:04:02.25 & -72:01:56.64 & 24.34 $\pm$ 0.03 & 23.93 $\pm$ 0.04 & 23.35 $\pm$ 0.03 & 0.41 $\pm$ 0.05 & 0.58 $\pm$ 0.05 & 2.92\arcsec \\ 
- & 01:04:03.04 & -72:01:55.95 & 23.29 $\pm$ 0.01 & 22.95 $\pm$ 0.02 & 22.48 $\pm$ 0.02 & 0.34 $\pm$ 0.02 & 0.47 $\pm$ 0.03 & 3.30\arcsec \\ 
- & 01:04:02.08 & -72:01:56.69 & 22.06 $\pm$ 0.01 & 21.87 $\pm$ 0.01 & 21.67 $\pm$ 0.01 & 0.18 $\pm$ 0.01 & 0.20 $\pm$ 0.01 & 3.34\arcsec \\ 
- & 01:04:01.95 & -72:01:51.68 & 21.11 $\pm$ 0.00 & 21.05 $\pm$ 0.01 & 20.93 $\pm$ 0.01 & 0.06 $\pm$ 0.01 & 0.12 $\pm$ 0.01 & 3.34\arcsec \\ 
- & 01:04:01.88 & -72:01:52.01 & 24.07 $\pm$ 0.02 & 23.81 $\pm$ 0.02 & 23.25 $\pm$ 0.02 & 0.26 $\pm$ 0.03 & 0.56 $\pm$ 0.03 & 3.39\arcsec \\ 
- & 01:04:03.25 & -72:01:54.82 & 21.65 $\pm$ 0.01 & 21.47 $\pm$ 0.01 & 21.22 $\pm$ 0.01 & 0.18 $\pm$ 0.01 & 0.25 $\pm$ 0.01 & 3.69\arcsec \\ 
106 & 01:04:01.66 & -72:01:54.55 & 20.42 $\pm$ 0.00 & 20.43 $\pm$ 0.01 & 20.45 $\pm$ 0.01 & -0.01 $\pm$ 0.01 & -0.02 $\pm$ 0.01 & 3.84\arcsec \\ 
A & 01:04:02.39 & -72:01:49.81 & 20.68 $\pm$ 0.00 & 20.69 $\pm$ 0.00 & 20.68 $\pm$ 0.00 & -0.01 $\pm$ 0.00 & 0.01 $\pm$ 0.00 & 4.13\arcsec \\ 
- & 01:04:03.19 & -72:01:51.35 & 23.01 $\pm$ 0.01 & 22.77 $\pm$ 0.01 & 22.37 $\pm$ 0.01 & 0.24 $\pm$ 0.01 & 0.40 $\pm$ 0.01 & 4.18\arcsec \\ 
- & 01:04:02.79 & -72:01:58.09 & 24.03 $\pm$ 0.02 & 23.54 $\pm$ 0.03 & 23.09 $\pm$ 0.02 & 0.49 $\pm$ 0.04 & 0.45 $\pm$ 0.04 & 4.42\arcsec \\ 
- & 01:04:01.76 & -72:01:56.92 & 22.66 $\pm$ 0.01 & 22.35 $\pm$ 0.02 & 21.95 $\pm$ 0.01 & 0.31 $\pm$ 0.02 & 0.40 $\pm$ 0.02 & 4.48\arcsec \\ 
- & 01:04:01.95 & -72:01:49.89 & 24.13 $\pm$ 0.02 & 23.74 $\pm$ 0.02 & 23.21 $\pm$ 0.02 & 0.39 $\pm$ 0.03 & 0.53 $\pm$ 0.03 & 4.71\arcsec \\ 
- & 01:04:03.49 & -72:01:53.18 & 23.89 $\pm$ 0.01 & 23.50 $\pm$ 0.02 & 22.99 $\pm$ 0.01 & 0.39 $\pm$ 0.02 & 0.51 $\pm$ 0.02 & 4.73\arcsec \\ 
- & 01:04:02.70 & -72:01:49.26 & 22.87 $\pm$ 0.01 & 22.56 $\pm$ 0.01 & 22.14 $\pm$ 0.01 & 0.31 $\pm$ 0.01 & 0.42 $\pm$ 0.01 & 4.77\arcsec \\ 
- & 01:04:03.21 & -72:01:50.25 & 22.97 $\pm$ 0.01 & 22.62 $\pm$ 0.01 & 22.11 $\pm$ 0.01 & 0.35 $\pm$ 0.01 & 0.52 $\pm$ 0.01 & 5.00\arcsec \\ 
- & 01:04:01.42 & -72:01:55.03 & 22.46 $\pm$ 0.01 & 22.22 $\pm$ 0.01 & 21.85 $\pm$ 0.01 & 0.24 $\pm$ 0.01 & 0.37 $\pm$ 0.01 & 5.05\arcsec \\ 
- & 01:04:03.63 & -72:01:52.32 & 22.18 $\pm$ 0.01 & 21.60 $\pm$ 0.01 & 20.78 $\pm$ 0.00 & 0.57 $\pm$ 0.01 & 0.82 $\pm$ 0.01 & 5.56\arcsec \\ 
- & 01:04:03.36 & -72:01:50.11 & 23.89 $\pm$ 0.01 & 23.53 $\pm$ 0.02 & 22.95 $\pm$ 0.01 & 0.35 $\pm$ 0.02 & 0.59 $\pm$ 0.02 & 5.59\arcsec \\ 
- & 01:04:03.19 & -72:01:58.46 & 21.26 $\pm$ 0.01 & 21.21 $\pm$ 0.01 & 21.19 $\pm$ 0.01 & 0.05 $\pm$ 0.01 & 0.02 $\pm$ 0.01 & 5.60\arcsec \\ 
B & 01:04:01.98 & -72:01:48.63 & 20.17 $\pm$ 0.00 & 19.47 $\pm$ 0.00 & 18.52 $\pm$ 0.00 & 0.70 $\pm$ 0.00 & 0.96 $\pm$ 0.00 & 5.78\arcsec \\ 
- & 01:04:02.32 & -72:01:59.94 & 24.36 $\pm$ 0.02 & 23.87 $\pm$ 0.03 & 23.16 $\pm$ 0.02 & 0.49 $\pm$ 0.04 & 0.70 $\pm$ 0.04 & 6.06\arcsec \\ 
- & 01:04:01.28 & -72:01:51.49 & 23.70 $\pm$ 0.01 & 23.32 $\pm$ 0.02 & 22.75 $\pm$ 0.01 & 0.38 $\pm$ 0.02 & 0.58 $\pm$ 0.02 & 6.07\arcsec \\ 
- & 01:04:01.90 & -72:01:59.41 & 24.46 $\pm$ 0.02 & 23.99 $\pm$ 0.03 & 23.44 $\pm$ 0.02 & 0.46 $\pm$ 0.04 & 0.55 $\pm$ 0.04 & 6.10\arcsec \\ 
- & 01:04:01.12 & -72:01:55.80 & 23.82 $\pm$ 0.02 & 23.49 $\pm$ 0.03 & 22.93 $\pm$ 0.02 & 0.33 $\pm$ 0.04 & 0.56 $\pm$ 0.04 & 6.58\arcsec \\ 
- & 01:04:01.18 & -72:01:51.20 & 23.23 $\pm$ 0.01 & 22.98 $\pm$ 0.01 & 22.48 $\pm$ 0.01 & 0.25 $\pm$ 0.01 & 0.50 $\pm$ 0.01 & 6.60\arcsec \\ 
- & 01:04:01.70 & -72:01:48.32 & 21.59 $\pm$ 0.00 & 21.43 $\pm$ 0.01 & 21.21 $\pm$ 0.01 & 0.16 $\pm$ 0.01 & 0.22 $\pm$ 0.01 & 6.66\arcsec \\ 
- & 01:04:03.91 & -72:01:52.78 & 22.96 $\pm$ 0.01 & 22.71 $\pm$ 0.01 & 22.17 $\pm$ 0.01 & 0.25 $\pm$ 0.01 & 0.54 $\pm$ 0.01 & 6.71\arcsec \\ 
- & 01:04:01.12 & -72:01:51.44 & 23.23 $\pm$ 0.01 & 22.93 $\pm$ 0.01 & 22.46 $\pm$ 0.01 & 0.30 $\pm$ 0.01 & 0.47 $\pm$ 0.01 & 6.78\arcsec \\ 
- & 01:04:01.50 & -72:01:48.74 & 24.21 $\pm$ 0.02 & 23.81 $\pm$ 0.02 & 23.26 $\pm$ 0.02 & 0.40 $\pm$ 0.03 & 0.55 $\pm$ 0.03 & 6.88\arcsec \\ 
6 & 01:04:03.04 & -72:02:00.30 & 19.81 $\pm$ 0.00 & 19.78 $\pm$ 0.00 & 19.79 $\pm$ 0.00 & 0.03 $\pm$ 0.00 & -0.01 $\pm$ 0.00 & 6.89\arcsec \\ 
- & 01:04:01.57 & -72:01:59.43 & 21.67 $\pm$ 0.00 & 21.52 $\pm$ 0.01 & 21.38 $\pm$ 0.01 & 0.16 $\pm$ 0.01 & 0.13 $\pm$ 0.01 & 6.94\arcsec \\ 
- & 01:04:03.71 & -72:01:49.84 & 23.68 $\pm$ 0.01 & 23.38 $\pm$ 0.02 & 22.89 $\pm$ 0.01 & 0.30 $\pm$ 0.02 & 0.49 $\pm$ 0.02 & 7.02\arcsec \\ 
- & 01:04:02.74 & -72:02:00.93 & 24.19 $\pm$ 0.02 & 23.73 $\pm$ 0.03 & 23.16 $\pm$ 0.02 & 0.46 $\pm$ 0.04 & 0.57 $\pm$ 0.04 & 7.11\arcsec \\ 
- & 01:04:00.92 & -72:01:53.61 & 21.56 $\pm$ 0.01 & 21.25 $\pm$ 0.01 & 20.94 $\pm$ 0.01 & 0.31 $\pm$ 0.01 & 0.31 $\pm$ 0.01 & 7.25\arcsec \\ 
- & 01:04:03.33 & -72:02:00.15 & 22.40 $\pm$ 0.01 & 22.05 $\pm$ 0.01 & 21.57 $\pm$ 0.01 & 0.35 $\pm$ 0.01 & 0.49 $\pm$ 0.01 & 7.37\arcsec \\ 
- & 01:04:03.80 & -72:01:49.77 & 22.50 $\pm$ 0.01 & 22.28 $\pm$ 0.01 & 21.96 $\pm$ 0.01 & 0.22 $\pm$ 0.01 & 0.32 $\pm$ 0.01 & 7.38\arcsec \\ 
144 & 01:04:00.96 & -72:01:56.38 & 20.53 $\pm$ 0.00 & 20.48 $\pm$ 0.01 & 20.48 $\pm$ 0.01 & 0.04 $\pm$ 0.01 & 0.01 $\pm$ 0.01 & 7.43\arcsec \\ 
- & 01:04:02.38 & -72:02:01.37 & 23.09 $\pm$ 0.01 & 22.66 $\pm$ 0.01 & 22.08 $\pm$ 0.01 & 0.43 $\pm$ 0.01 & 0.57 $\pm$ 0.01 & 7.46\arcsec \\ 
- & 01:04:00.93 & -72:01:56.13 & 24.40 $\pm$ 0.03 & 23.88 $\pm$ 0.04 & 23.40 $\pm$ 0.03 & 0.52 $\pm$ 0.05 & 0.49 $\pm$ 0.05 & 7.49\arcsec \\ 
- & 01:04:01.75 & -72:01:47.19 & 23.55 $\pm$ 0.01 & 23.28 $\pm$ 0.02 & 22.79 $\pm$ 0.01 & 0.28 $\pm$ 0.02 & 0.49 $\pm$ 0.02 & 7.53\arcsec \\ 
- & 01:04:01.08 & -72:01:49.92 & 22.81 $\pm$ 0.01 & 22.53 $\pm$ 0.01 & 22.10 $\pm$ 0.01 & 0.28 $\pm$ 0.01 & 0.44 $\pm$ 0.01 & 7.63\arcsec \\ 
- & 01:04:03.47 & -72:02:00.10 & 23.22 $\pm$ 0.01 & 22.89 $\pm$ 0.01 & 22.42 $\pm$ 0.01 & 0.33 $\pm$ 0.01 & 0.48 $\pm$ 0.01 & 7.70\arcsec \\ 
- & 01:04:00.90 & -72:01:56.42 & 23.23 $\pm$ 0.01 & 22.92 $\pm$ 0.02 & 22.48 $\pm$ 0.02 & 0.30 $\pm$ 0.02 & 0.44 $\pm$ 0.03 & 7.74\arcsec \\ 
- & 01:04:00.81 & -72:01:52.86 & 22.97 $\pm$ 0.01 & 22.68 $\pm$ 0.02 & 22.32 $\pm$ 0.01 & 0.29 $\pm$ 0.02 & 0.36 $\pm$ 0.02 & 7.79\arcsec \\ 
- & 01:04:03.88 & -72:01:49.44 & 24.15 $\pm$ 0.02 & 23.84 $\pm$ 0.03 & 23.23 $\pm$ 0.02 & 0.31 $\pm$ 0.04 & 0.61 $\pm$ 0.04 & 7.90\arcsec \\ 
- & 01:04:01.09 & -72:01:49.07 & 22.04 $\pm$ 0.01 & 21.86 $\pm$ 0.01 & 21.62 $\pm$ 0.01 & 0.18 $\pm$ 0.01 & 0.24 $\pm$ 0.01 & 8.04\arcsec \\ 
- & 01:04:03.22 & -72:02:01.21 & 23.68 $\pm$ 0.01 & 23.36 $\pm$ 0.02 & 22.79 $\pm$ 0.01 & 0.32 $\pm$ 0.02 & 0.57 $\pm$ 0.02 & 8.06\arcsec \\ 
- & 01:04:01.23 & -72:01:59.59 & 24.14 $\pm$ 0.02 & 23.74 $\pm$ 0.02 & 23.16 $\pm$ 0.02 & 0.41 $\pm$ 0.03 & 0.57 $\pm$ 0.03 & 8.11\arcsec \\ 
- & 01:04:01.36 & -72:02:00.17 & 24.77 $\pm$ 0.03 & 24.02 $\pm$ 0.04 & 23.12 $\pm$ 0.02 & 0.75 $\pm$ 0.05 & 0.90 $\pm$ 0.04 & 8.12\arcsec \\ 
- & 01:04:04.03 & -72:01:57.76 & 21.83 $\pm$ 0.01 & 21.68 $\pm$ 0.01 & 21.45 $\pm$ 0.01 & 0.16 $\pm$ 0.01 & 0.23 $\pm$ 0.01 & 8.14\arcsec \\ 
- & 01:04:03.44 & -72:01:46.75 & 23.34 $\pm$ 0.01 & 22.99 $\pm$ 0.02 & 22.48 $\pm$ 0.01 & 0.34 $\pm$ 0.02 & 0.51 $\pm$ 0.02 & 8.44\arcsec \\ 
- & 01:04:00.66 & -72:01:55.02 & 24.45 $\pm$ 0.03 & 24.00 $\pm$ 0.04 & 23.43 $\pm$ 0.03 & 0.45 $\pm$ 0.05 & 0.57 $\pm$ 0.05 & 8.48\arcsec \\ 
- & 01:04:02.86 & -72:01:45.58 & 23.92 $\pm$ 0.01 & 23.53 $\pm$ 0.02 & 23.00 $\pm$ 0.01 & 0.39 $\pm$ 0.02 & 0.52 $\pm$ 0.02 & 8.52\arcsec \\ 
- & 01:04:04.34 & -72:01:54.39 & 22.65 $\pm$ 0.01 & 22.36 $\pm$ 0.01 & 21.89 $\pm$ 0.01 & 0.28 $\pm$ 0.01 & 0.47 $\pm$ 0.01 & 8.61\arcsec \\ 
- & 01:04:04.12 & -72:01:49.68 & 23.09 $\pm$ 0.01 & 22.78 $\pm$ 0.01 & 22.26 $\pm$ 0.01 & 0.31 $\pm$ 0.01 & 0.52 $\pm$ 0.01 & 8.68\arcsec \\ 
- & 01:04:03.00 & -72:01:45.55 & 21.27 $\pm$ 0.00 & 21.17 $\pm$ 0.01 & 21.04 $\pm$ 0.01 & 0.10 $\pm$ 0.01 & 0.13 $\pm$ 0.01 & 8.70\arcsec \\ 
- & 01:04:00.59 & -72:01:54.70 & 22.70 $\pm$ 0.01 & 22.42 $\pm$ 0.02 & 21.97 $\pm$ 0.01 & 0.28 $\pm$ 0.02 & 0.45 $\pm$ 0.02 & 8.77\arcsec \\ 
C & 01:04:03.69 & -72:01:47.14 & 20.86 $\pm$ 0.00 & 20.85 $\pm$ 0.01 & 20.84 $\pm$ 0.00 & 0.01 $\pm$ 0.01 & 0.00 $\pm$ 0.01 & 8.79\arcsec \\ 
- & 01:04:02.47 & -72:02:02.73 & 23.09 $\pm$ 0.01 & 22.67 $\pm$ 0.01 & 22.25 $\pm$ 0.01 & 0.43 $\pm$ 0.01 & 0.42 $\pm$ 0.01 & 8.81\arcsec \\ 
- & 01:04:01.72 & -72:01:45.78 & 23.61 $\pm$ 0.01 & 23.22 $\pm$ 0.02 & 22.65 $\pm$ 0.01 & 0.38 $\pm$ 0.02 & 0.58 $\pm$ 0.02 & 8.87\arcsec \\ 
- & 01:04:02.14 & -72:02:02.67 & 23.19 $\pm$ 0.01 & 22.80 $\pm$ 0.01 & 22.40 $\pm$ 0.01 & 0.38 $\pm$ 0.01 & 0.41 $\pm$ 0.01 & 8.89\arcsec 
\enddata
\label{table:photometryNew}
\end{deluxetable*}

\begin{figure*}[h]  
\begin{center}
\hspace{-0.5cm}
\includegraphics[width=40pc]{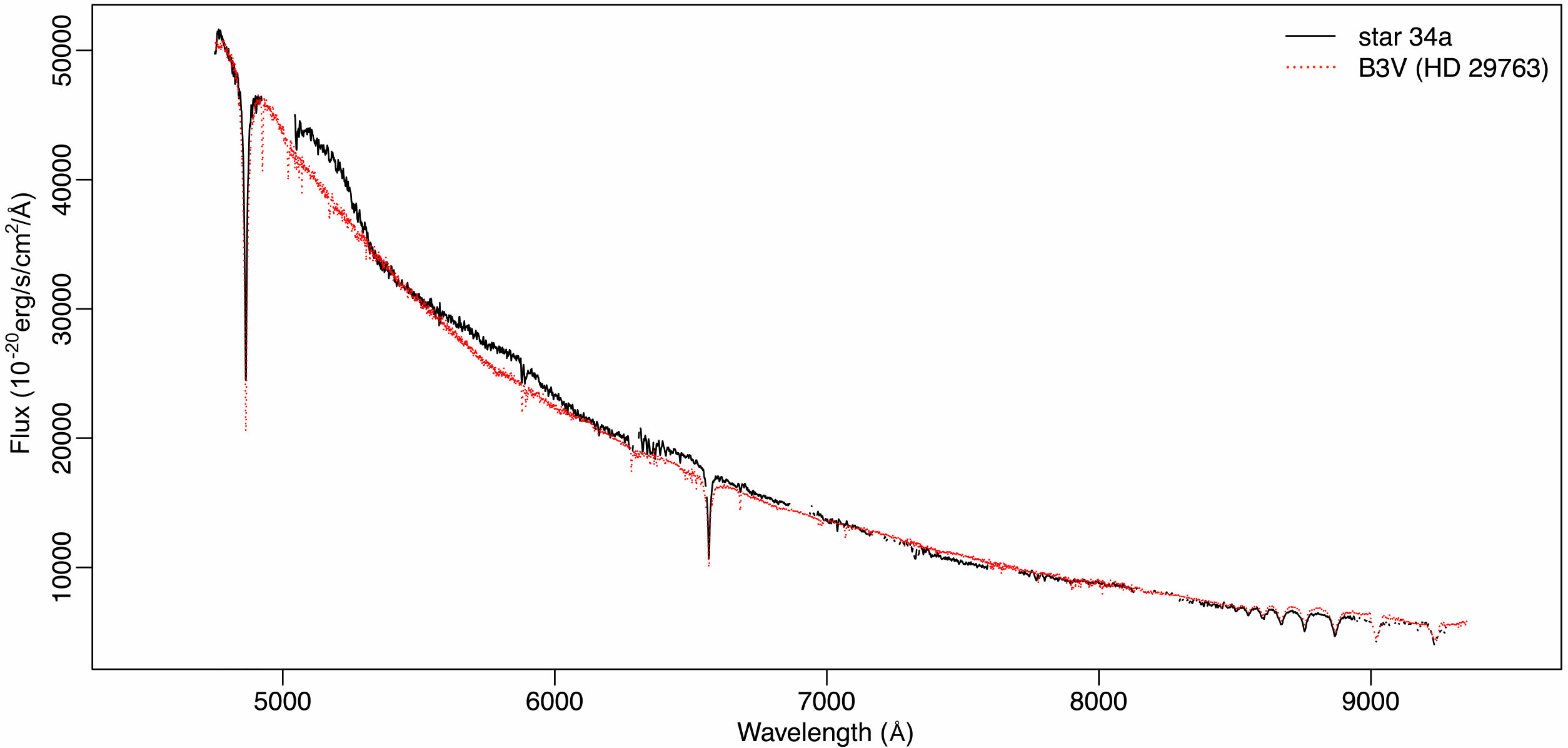}
\caption{
The VLT MUSE dereddened spectrum of star 34a and a template spectrum taken from the Indo-US Library of Coud$\grave{e}$ Feed Stellar Spectra \citep{Valdes2004}. The scale of the template spectrum is adjusted to the observed spectrum.}
\label{figure:Star34a}
\end{center}
\end{figure*}

\begin{figure*}[h]  
\begin{center}
\hspace{-0.5cm}
\includegraphics[width=40pc]{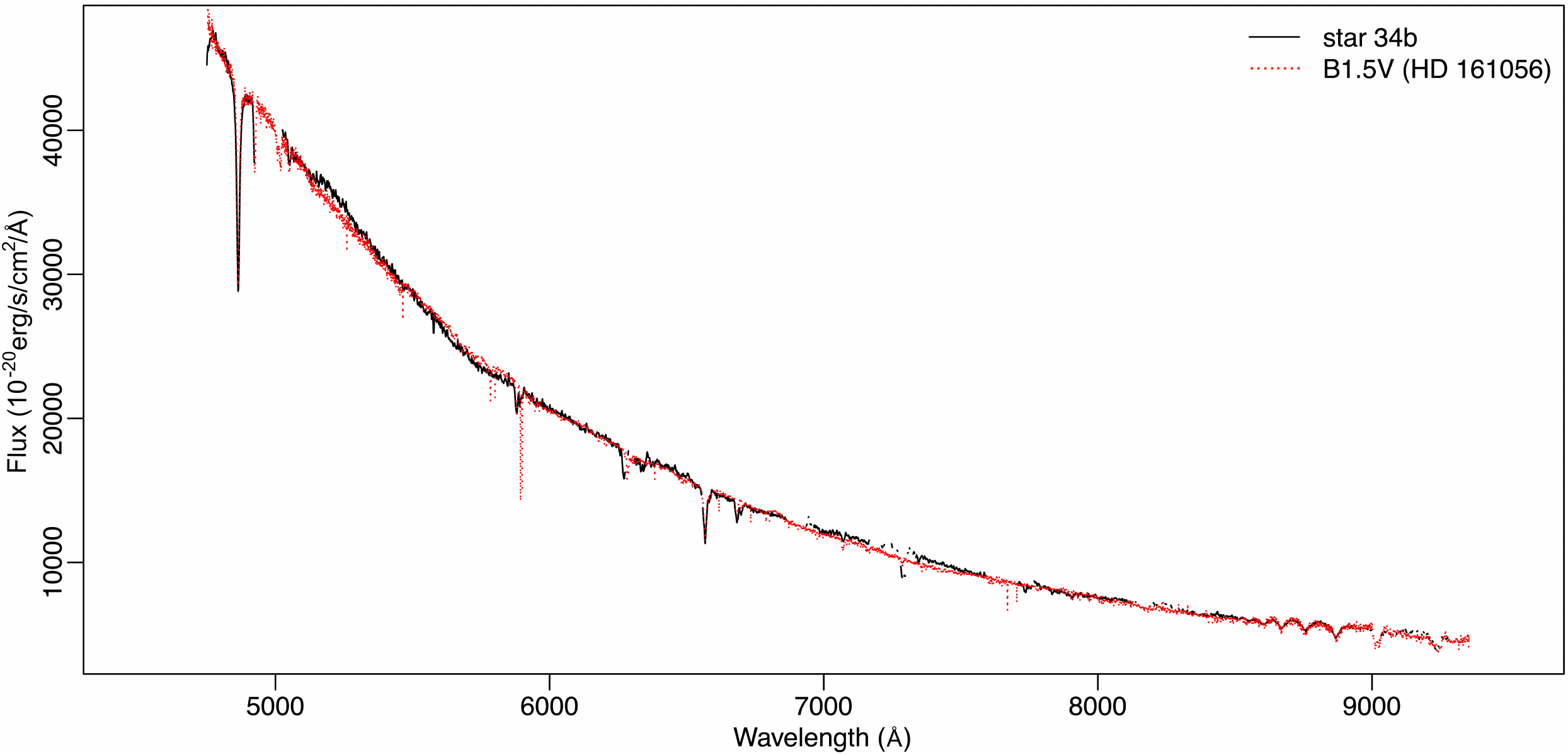}
\caption{ 
Same as Figure \ref{figure:Star34a}, but for star 34b.
}
\label{figure:Star34b}
\end{center}
\end{figure*}

\end{document}